\documentclass[fleqn,usenatbib]{mnras_nomnras}
\usepackage{newtxtext,newtxmath}
\usepackage[T1]{fontenc}
\usepackage{float}

\DeclareRobustCommand{\VAN}[3]{#2}
\let\VANthebibliography\thebibliography
\def\thebibliography{\DeclareRobustCommand{\VAN}[3]{##3}\VANthebibliography}

\usepackage{CJKutf8}
\usepackage{booktabs}
\usepackage{xcolor, soul}
\usepackage{amsmath,bm} % amssymb,
\usepackage{graphicx}
\title[Guide to the scattering transform]{How to quantify fields or textures? A guide to the scattering transform}

\author[Cheng \& M\'enard]{
Sihao Cheng (程思浩)$^{1,2}$\thanks{E-mail: s.cheng@jhu.edu} \&
Brice M\'enard$^{1,2}$
\\
$^{1}$Department of Physics and Astronomy, The Johns Hopkins University, 3400 N Charles Street, Baltimore, MD 21218, USA\\
$^{2}$Centre de Sciences des Donn\'ees, Ecole Normale Sup\'erieure, 45 Rue d'Ulm, 75005, Paris, France
}

\date{\today}
\pubyear{2021}

\begin{document}
\label{firstpage}
\pagerange{\pageref{firstpage}--\pageref{lastpage}}
\begin{CJK}{UTF8}{gkai} 
\maketitle
\end{CJK}

\begin{abstract}
Extracting information from stochastic fields or textures is a ubiquitous task in science, from exploratory data analysis to classification and parameter estimation. From physics to biology, it tends to be done either through a power spectrum analysis, which is often too limited, or the use of convolutional neural networks (CNNs), which require large training sets and lack interpretability. In this paper, we advocate for the use of the scattering transform \citep{Mallat_2012}, a powerful statistic which borrows mathematical ideas from CNNs but does not require any training, and is interpretable. We show that it provides a relatively compact set of summary statistics with visual interpretation and which carries most of the relevant information in a wide range of scientific applications. We present a non-technical introduction to this estimator and we argue that it can benefit data analysis, comparison to models and parameter inference in many fields of science. Interestingly, understanding the core operations of the scattering transform allows one to decipher many key aspects of the inner workings of CNNs.
\end{abstract}


% We need to say why complex wavelets.

\section{Introduction}
\label{sec:intro}

To understand the laws of Nature, most physicists face the challenge of extracting relevant information from data produced by sensors or computer simulations. Different sets of data analysis tools are used in different disciplines, sometimes motivated by the properties of the object of study (symmetries, invariants), aspects of the data (noise level) or simply convenience or habit. Among them, a few are used ubiquitously, such as the power spectrum and correlation functions. Over the past decade, a new type of estimator has gained popularity in virtually all fields of science: convolutional neural networks (CNNs), a novel paradigm remarkably efficient at extracting certain types of information from pixelized data but whose properties are not yet fully understood. Compared to the traditional mathematical tools, neural networks lack transparency, stable mathematical properties or interpretability, which are crucial to scientific research.

With growing depth, neural networks enjoy a fantastic level of expressivity, capable of capturing the highly complex sets of varying features produced by the biological world, from the appearance of cats to human speech. However, when considering a wide range of physical fields, we are in a different regime. We often deal with a level of complexity substantially lower than that involved in the typical images or sounds considered in deep learning applications. A question then arises: which estimator should be considered? In this paper, we advocate for the use of the \emph{scattering transform}, introduced by \citet{Mallat_2012} to extract information from physical fields. It provides an approach to data analysis that in many ways conveniently stands \emph{in between} the power spectrum and CNNs. The scattering transform has many attractive properties. It can efficiently extract information from complex signals while being fully deterministic and not requiring any training. In addition, understanding the properties of the scattering transform allows one to decipher many key aspects of the inner workings of CNNs. 

The scattering transform was originally introduced in the mathematics literature with follow-up works that appeared in the signal processing and computer science literature. So far, it has been used primarily in audio/visual signal processing
\citep[e.g.,][]{AndenMallat_2011, Bruna_2013, Sifre_2013, AndenMallat_2014}.
It has already been used in a number of scientific applications:
intermittency in turbulence \citep{Bruna_2015},
quantum chemistry and material science \citep{Hirn_2017, Eickenberg_2018, Sinz_2020}, plasma physics \citep{Glinsky_2020},
% graph-structured data \citep{gama2018diffusion},
geography \citep{Kavalerov_2019},
astrophysics \citep{Allys_2019,Saydjari_2021, Blancard_2020}, 
and cosmology \citep{Cheng_2020, Cheng_2021}.
%Electrocardiograms
%Balestriero, R., Cosentino, R., Glotin, H. & Baraniuk, R. Spline filters for end-to- end deep learning. in Proceedings of the 35th International Conference on Machine Learning, Vol. 80 of Proceedings of Machine Learning Research. (eds Dy, J. & Krause, A.) 364–373 (PMLR, Stockholmsmässan, Stockholm, Sweden, 2018).
In several of these applications, the scattering transform reached state-of-the-art performance compared to the CNNs in use at the time. It has been typically used in the context of classification tasks but, as we will discuss, it is  powerful for a wide range of applications, from exploratory data analysis to regression or parameter inference when models are available.

We believe that scientists in many disciplines can  benefit from adoption the adoption of the scattering transform in their research. In this paper, we present it in a ``non-technical way''. Our presentation differs from that of the original papers in the mathematics literature: we focus mainly on the aspects directly relevant to the analysis of finite datasets and we de-emphasise or omit properties potentially interesting to mathematicians (such as behaviours at infinity) but with limited applicability to actual data. After summarising the key features of the scattering transform, we discuss the interpretation of its coefficients and the role of its internal operations.

\begin{figure*}
    \centering
    \includegraphics[width=\textwidth]{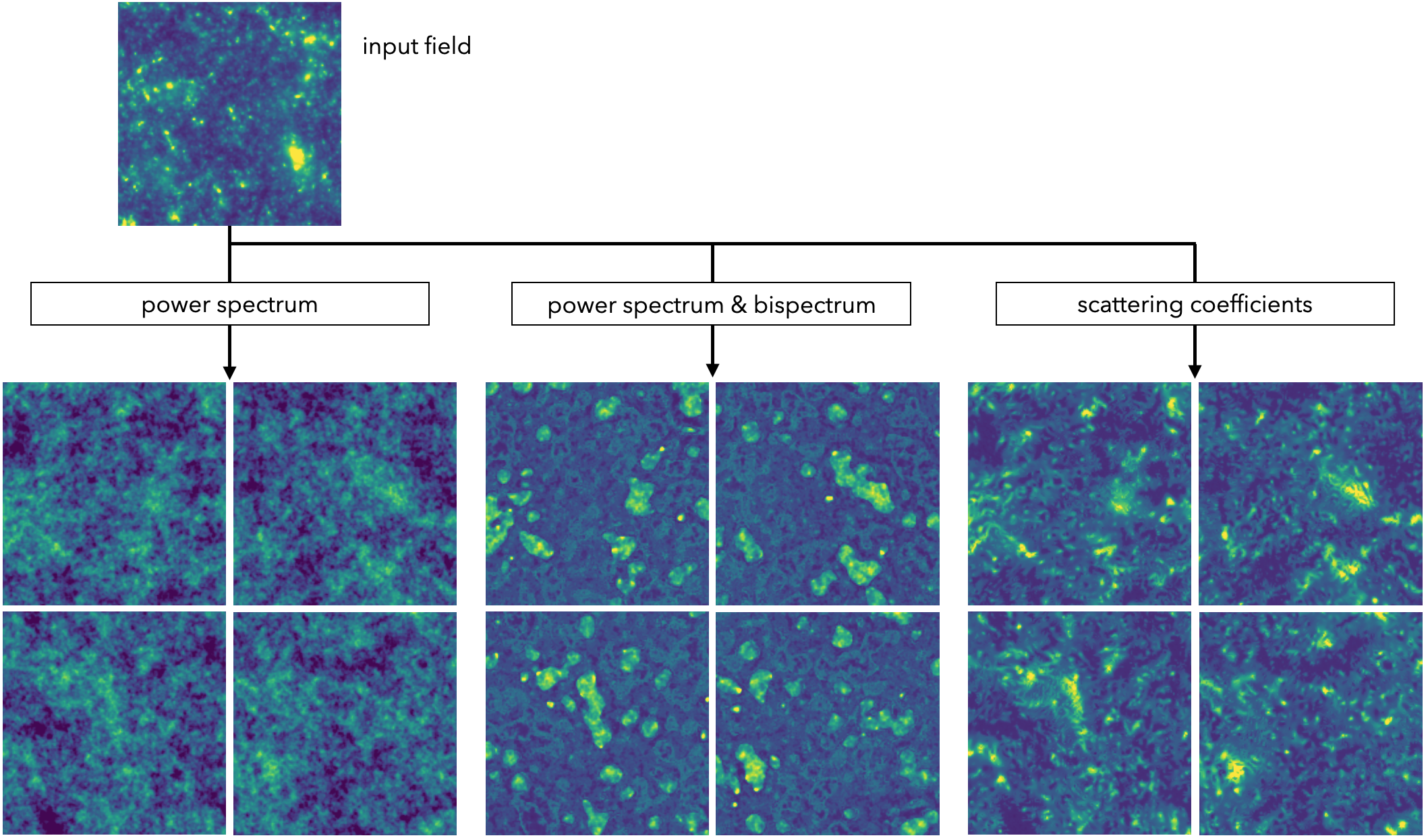}
    \caption{Texture synthesis performed using different translation-invariant summary statistics. In each case, the four synthesized fields are obtained using four sets of initial random fluctuations gradually evolved to match the summary statistics of the input field. As can be seen, the scattering transform performs significantly better. Detailed algorithm is described in appendix~\ref{app:synthesis}.}
    \label{fig:PS_limitation}
\end{figure*}

\section{Extracting information from a field}

\subsection{Goals and challenges}
\label{sec:characterisation}

Extracting physical information from data typically requires a mapping from the extremely high-dimensional function space of data to a low-dimensional space that corresponds to a limited number of classes or parameters. To do so, 
it is convenient and often necessary to first describe the field with a mathematical vocabulary, which aims at discarding irrelevant variabilities and concentrating relevant information into a smaller set of descriptors or summary statistics. Finding the best language or description to extract information from data is often the key challenge of data analysis.

When considering physical systems, a number of properties are fundamental. For example, understanding how energy is distributed is often an important goal. In addition, symmetries and invariants often play a major role. Many processes are invariant under translation in space and/or time. When relevant, such properties should be directly incorporated into the chosen representation to discard the irrelevant variability. Ideally, the summary statistic should also be robust and compact, which are necessary for interpretability. Robustness requires stability or continuity of the descriptors with respect to potential perturbations of the field, such as additive noise, geometric or temporal deformation, 
change of parameters of the field, 
potential distortions from the detector, etc.
Within the community of computer vision, a lot of attention has been paid to classification problems and the critical need for stable estimators. We point out that, despite being often less discussed or even omitted in 
the context of scientific applications, the need for stability is as important for exploratory data analysis and parameter inference (regression).

In summary, to extract information from a field, the challenge is often reduced to finding a set of (statistical) descriptors that are invariant to translation and/or rotation, robust, and compact, while being informative. To better understand how to obtain such properties in a statistical estimator, we begin by discussing the key aspects and limitations of the most commonly used summary statistic in scientific applications: the power spectrum. In order to simplify the formalism and discussion, we restrict our domain to stationary ergodic processes. In other words, we will primarily characterize \emph{textures}. Extensions to \emph{objects} will be discussed towards the end.

\begin{figure*}
    \includegraphics[width=\textwidth]{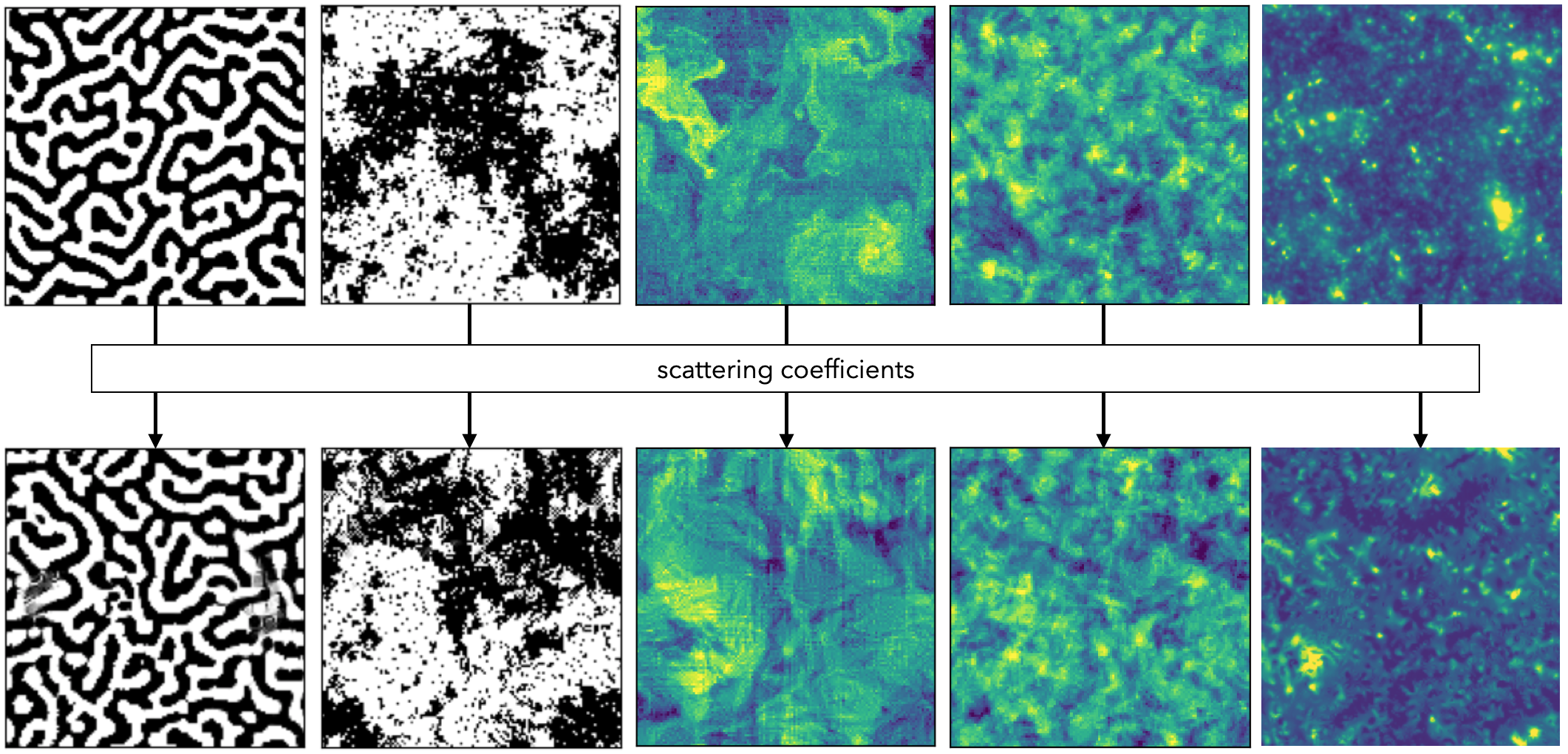}
    \caption{
    Texture synthesis using the scattering transform for a variety of physical fields: Turing pattern, Ising model, ocean turbulence, solar surface, cosmic matter density. The upper panels show input 2-D fields from simulations or observations. The lower panels show randomly generated fields with scattering coefficients matching their upper counterparts.
    }
    \label{fig:synthesis}
\end{figure*}

\subsection{The ubiquitous power spectrum}
\label{sec:PS}

The power spectrum, also called the spectral density, spectral power density, or energy spectral density, is a ubiquitous statistical descriptor in natural sciences. It is used in all kinds of studies, from the most theoretical works to exploratory data analyses. The power spectrum quantifies the variance of a field as a function of frequency, usually temporal or spatial. In many cases, it is related to the physical energy (or energy density) of the system: for example, for fluids, the variance of velocity field is the kinetic energy; for fields described by wave equations, the variance corresponds to the potential energy. The power spectrum is often a useful tool to extract characteristic scales in a system or a field.

Given a field $I(\vec x)$, the power spectrum is defined through its Fourier transform, $\tilde I(\vec k)$, as
\begin{equation}
\label{eq:P(k)}
P(\vec k) \equiv \tilde{I}(\vec k)\tilde{I}(-\vec k)=|\tilde{I}(\vec k)|^2 \;,
\end{equation}
where $\vec k$ is a frequency.
Its generic use is motivated by a number of properties:
\begin{itemize}
    \item \textbf{Translation-invariance:} 
    % The Fourier transform is sensitive to translations but its modulus is not. 
    being only a function of Fourier amplitudes, the power spectrum is 
    % Defined from Fourier modulus, the power spectrum is translation-invariant and therefore 
    immune to irrelevant variability introduced by spatial or temporal shifts.

    \item \textbf{Energy extraction and scale separation:} 
    The power spectrum is a partition of the variance or energy density of the field as a function of scale, due to the orthogonality of Fourier modes. This is often an important quantity to extract characteristic frequency/scales of a system, the range of scales above/below which noise contributions dominate and its decay rate as a function of frequency informs on the regularity of the fluctuations. 
    
    \item \textbf{Dimensionality reduction:}
    The power spectrum itself does not compress data, but two common binning schemes can substantially reduce the number of coefficients without losing much information. 1) Statistical isotropy allows for an average over all orientations. 2) smoothness of the expected power spectrum, (equivalent to that the correlation decays fast enough that at long distance) enables a binning in neighbouring scales.
    
    \item \textbf{Asymptotic normality:} 
    due to the central limit theorem, the amplitudes of the power spectrum measured in bins of frequencies are subject to Gaussianization. This is a desirable property for a summary statistic. Indeed, being able to model the sampling distribution by a Gaussian facilitates the parameterization of the likelihood function needed for parameter inference.
    
    \item \textbf{Theoretical predictions:}
    the mathematics of power spectrum are familiar to theorists. When the field is in a perturbative regime, the power spectrum can be accurately predicted from the field equation. 
\end{itemize}

\begin{figure*}
    \centering
    \includegraphics[width=0.7\textwidth]{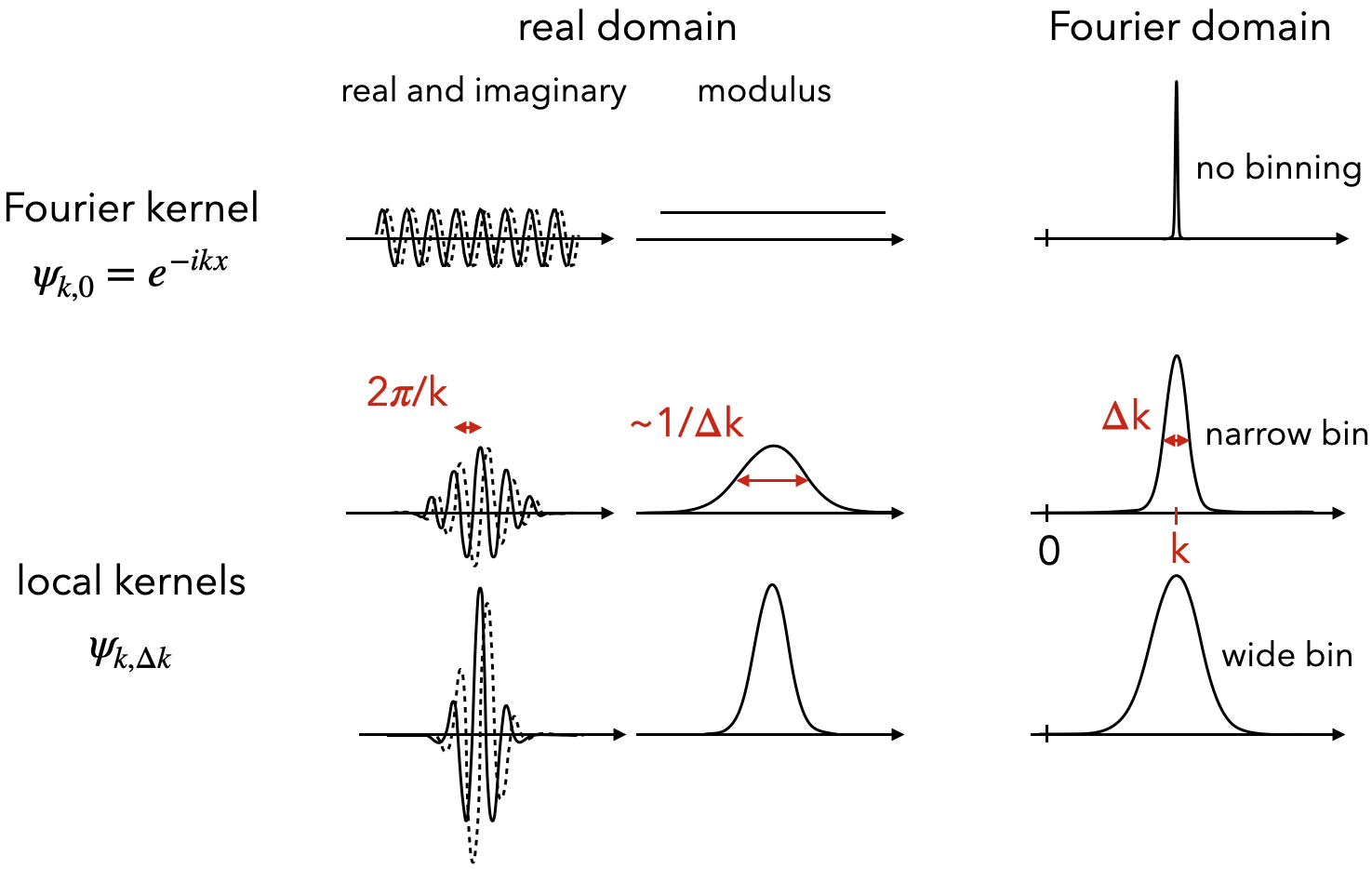}
    \caption{Comparison of Fourier and wavelet kernels. The Fourier kernels are fully localized in frequency space and delocalized in real space. Instead, the wavelets are partially localized in both domains.}
    \label{fig:Fourier_vs_wavelet}
\end{figure*}

However, power spectrum-based analyses have a serious limitation: while it is the adequate estimator to extract the energy, the power spectrum is usually insufficient for information extraction as it only estimates the variance of fluctuations as a function of scale. This is equivalent to using a ellipsoid ball to fit the distribution function of the  random field, which is sufficient only for Gaussian random fields. However, the vast majority of fields are not in this restricted regime: they display so-called non-Gaussianities characterized by interactions between scales which generate specific morphological features in the fields. This non-Gaussian information is carried largely by the Fourier phases of the field, which do not contribute to the power spectrum. This is illustrated in figure~\ref{fig:PS_limitation}: the power spectrum of an input field is measured and then used to generate new fields by evolving random fluctuations until they possess the same power spectrum. As can be seen, while the new images have the same second-order statistics, all morphological information has been lost.

Additional information can be extracted by measuring higher-order statistics of the field, as illustrated in figure~\ref{fig:PS_limitation}. Unfortunately, this approach is plagued with two main issues:
\begin{itemize}
    \item non-robustness/divergence: higher-order statistics are obtained by \emph{multiplying} combinations of random variables. This process amplifies the variability of the input signal, increasing its variance as well as all higher-order moments, which reduces the convergence rate towards asymptotic normality, causing non-robustness and, in some cases, leading to divergence.
    \item information dilution: the number of coefficients required to describe the possible configurations of higher-order statistics increases steeply with the order $n$, leading to a diluted description of the relevant information.
\end{itemize}
These issues make higher-order statistics inefficient in concentrating information and hard to use in practice. We will further discuss these points in section~\ref{sec:higher-order}. Before formally introducing the scattering transform as a way to extract information beyond the power spectrum, we present two lines of intuition leading to its key properties: (i) by re-orienting the power spectrum approach and (ii) by simplifying the principle behind a CNN.

\subsection{First intuition: expanding the power spectrum approach}
\label{sec:P_to_S}

In order to design an estimator more informative than the power spectrum while avoiding the drawbacks of higher-order statistics, it is instructive to examine the formalism of the power spectrum in real space. Instead of being defined as the two-point multiplications in Fourier space (eq.~\ref{eq:P(k)}), the power spectrum can be equivalently calculated in real space as a spatial average:
\begin{equation}
    P(k) \propto \langle |I \star \psi_k'|^2 \rangle\,,
\label{eq:P}
\end{equation}
where $\psi_k'$ is the Fourier kernel e$^{-ikx}$. From this point of view, the power spectrum involves (i) convolutions by a series of kernels, (ii) a point-wise non-linear function (the squared modulus), and (iii) an average. This series of operations remind us of the key operations used in CNNs which also make use of (i) a sequence of convolutions with localised kernels learned during a training phase followed by (ii) a non-linear function and (iii) a `pooling' operation, i.e., an average or a max of neighbouring pixels. Therefore, \emph{the power spectrum can be seen as a 1-layer CNN with pre-determined kernels}. Computing a power spectrum is thus similar to forward-propagating a trained CNN.

The real-space expression of the power spectrum in eq.~\ref{eq:P} can be interpreted as follows: the convolution \emph{selects} fluctuations/features at a given scale; the non-linearity (modulus square) estimates the strength of the fluctuations/features; and the average extracts the global value over the field. If we do not restrict ourselves to two-point statistics, two modifications of the power spectrum can be considered, based on its real-space interpretation and its connection to CNNs. First, we can replace the delocalized Fourier kernels $\psi'_k=e^{-ikx}$ by localized ones $\psi_k$, for example a family of wavelets. Next, we point out that squaring the modulus is necessary for converting fluctuations into their strength. Instead, we can simply use the modulus to obtain a \emph{lower-order} statistic
\begin{align}
    S_1(k) \equiv \langle | I \star \psi_k | \rangle\,,
\end{align}
which is qualitatively similar to the power spectrum. 

Interestingly, this new approach allows us to probe scale interactions by simply re-applying the operation
\begin{equation}
    S_2(k_1, k_2) \equiv \langle | |I \star \psi_{k_1} | \star \psi_{k_2} | \rangle\;.
\end{equation}
This would not be possible with the Fourier kernels $\psi'$ used in the power spectrum. The locality of kernels here is crucial.
As illustrated in figure~\ref{fig:Fourier_vs_wavelet}, if the kernels are the delocalized Fourier modes, 
all local information will be lost and it is no longer possible to probe interactions between scales through this approach. Also, dropping the square of the modulus ensures that all the $S_n$ statistics are lower-order statistics, which are more robust to additive perturbations of the field.

The $S_1$ and $S_2$ coefficients are the first and second-order scattering coefficients which will be introduced more generically in section~\ref{sec:formalism}. They can characterise some of the properties of a field beyond the limited Gaussian information described by the power spectrum. In particular, the scattering coefficients are sensitive to sparsity and interactions among scales. We also point out that this proposed approach differs from the usual higher-order moments for which the power index of the input field is elevated. The scattering transform approach instead uses successive applications of the convolution and modulus, allowing the estimator to stay `first-order' in the input data.

\begin{figure*}
    \centering
    \includegraphics[width=0.7\textwidth]{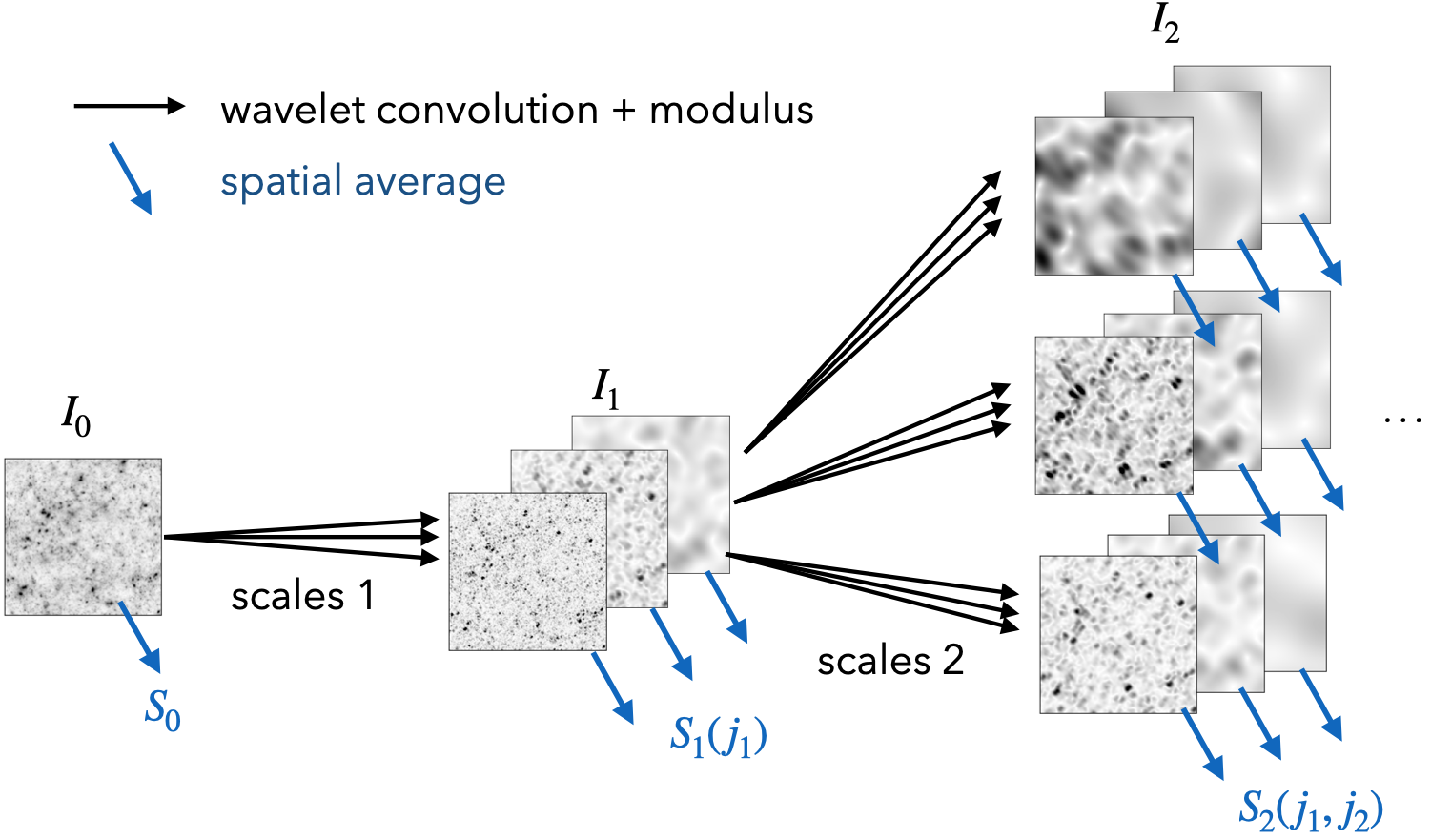}
    \caption{An illustration of a second-order scattering transform considering three scales. 
    The corresponding structure resembles the multi-layer architecture of a convolutional neural network (CNN).
    However, the scattering transform uses pre-determined wavelets and does not require any training.
    }
    \label{fig:ST_tree}
\end{figure*}

\subsection{Second intuition: simplifying a CNN}

Additional insights into the design of the scattering transform can be obtained by simplifying certain aspects of a convolutional neural network. The main modification is to use a pre-determined set of convolution kernels instead of learning them from the training set through challenging and costly optimisation. 

As pointed out by \citet{Bruna_2013}, the key elements found in the architecture of CNNs can be simplified or replaced as follows:
\begin{itemize}
    \item learnable convolutions $\rightarrow$ pre-defined wavelet convolutions
    \item ReLU $\rightarrow$ modulus
    \item pooling $\rightarrow$ average
    \item multi-layer $\rightarrow$ iterative approach
\end{itemize}
With those simplifications in mind, the forward-propagation operation done in each layer of a CNN can be written as the `scattering operation'. For example, the first layer becomes
\begin{align}
    I_0\rightarrow I_1\equiv|I_0 \star \psi|
\end{align}
where $\psi$ is a set of pre-determined convolution kernels. An activation amplitude over the entire field can then be obtained by $S_1=\langle I_1 \rangle$.
The connection of two layers extracting information from two different scales can be obtained by the
successive application of the scattering operation for two kernels
\begin{align}
    I_2 \equiv | |I_0 \star \psi_{1} | \star \psi_{2} | 
\end{align} 
and an activation amplitude over the entire field can similarly be obtained by $S_2=\langle I_2 \rangle$.
Generalizing this to $n$ layers:
\begin{align}
    I_n \equiv | |I_0 \star \psi_{1} | \star \psi_{2} |  ... \star \psi_{n} | \,,
\end{align}
results in a convolution tree or planar network which outputs a set of coefficients $S_n= \langle I_n \rangle$, the scattering coefficients. Forward-propagation into such a simplified CNN is therefore similar to calculating scattering coefficients. With the insight provided by both the expansion of the power spectrum and the simplification of CNNs, we now present the full formalism of the scattering transform.

\section{The scattering transform}
\label{sec:formalism}

The scattering transform \citep{Mallat_2012, Bruna_2013} was originally developed in the context of signal processing in computer vision. It has implicit connections to the power spectrum and CNNs, and it shares advantages from both sides. It is mathematically well-defined, interpretable and can perform remarkably well in the statistical extraction of information. Here, we present a condensed version of its formalism focusing only on the properties potentially relevant for typical scientific data analyses. For the full mathematical motivation, construction and properties of the estimator under infinite expansions, we refer the reader to \citep[see, e.g.,][]{Mallat_2012, Bruna_2013, Sifre_2013, AndenMallat_2014, Bruna_2015}.

\subsection{Formalism}

Conceptually, the scattering transform is composed of wavelet convolutions, modulus, hierarchy, and average. It yields translation-invariant descriptors $S_n(j_1, ..., j_n)$ from an input field $I_0(x)$ by recursively applying the following operations:
\begin{align}
    I_{n-1} \rightarrow I_n &\equiv \left|I_{n-1} \star \psi^{j}\right|\,\text{ (the scattering operation)}\label{eq:I_n}\\
    S_n &\equiv \langle I_n \rangle\,,
\end{align}
where $\psi^{j}$ stands for a wavelet indexed by $j$, its logarithmic scale. When the signal $I(\vec{x})$ is higher than one dimension, an index $l$ for the orientation of wavelet should also be added: $j \rightarrow j, l$. Considering stationary ergodic processes, the ensemble average can be estimated from an average over the extent $x$ of a realization. 
 
As illustrated in figure~\ref{fig:ST_tree}, successive applications of the scattering operation form a tree structure, i.e. a planar multi-layer network, with the scattering fields $I_n(x)$ at its nodes. Each $I_n$ is the intensity map of around a scale in the previous-order field $I_{n-1}$, which is similar to a local power spectrum analysis of the previous-order field. The average operation at each node  is used to extract a translation-invariant scattering coefficient. It is similar to the pooling operation in convolutional neural networks. The 0th-, 1st-, and 2nd-order scattering coefficients can be written explicitly as:
\begin{align}
    S_0 &\equiv \langle I_0 \rangle \label{eq:S0}\\
    S_1(j_1) &\equiv \langle I_1^{j_1}~~~~~~~\rangle = \langle |I_0\star\psi^{j_1}| \rangle \label{eq:S1}\\
    S_2(j_1,j_2) &\equiv \langle I_2^{j_1,j_2}\rangle = \langle  \left| |I_0\star\psi^{j_1}|\star\psi^{j_2}\right| \rangle\,. \label{eq:S2}
\end{align}
As mentioned before, when $I_0(\vec{x})$ is higher than one dimension, such as an image, the scale index $j$ becomes a scale index $j$ and an orientation index $l$. The kernels $\psi^{j}$ (or $\psi^{j,l}$) are chosen to be a family of wavelets, which are produced by dilating and rotating a mother wavelet.

\subsubsection*{Number of scattering coefficients}

The number of scattering coefficients is determined by the number of wavelet combinations. It is usual to consider a dyadic sequence of scales: $2^j$, with integer $1 \le j \le J$ which cannot exceed the length of the signal $2^J$. With $J$ choices of scales for each wavelet, there are $J^n$ available combinations at the $n$th order. However, only a subset carries relevant information. As we will explain in section~\ref{sec:translation_invariant}, the modulus operation extracts the envelope of the signal which scatters information and energy only into larger scales. As a result, only combinations with $j_2>j_1$ are significant, which reduces the number of informative coefficients by a factor of $2^{n-1}$. Thus, the number of useful scattering coefficients at each order is:
\begin{align*}
    &\text{0\textsuperscript{th} order: 1 coefficient,}\\
    &\text{1\textsuperscript{st} order: $J$ coefficients,}\\
    &\text{2\textsuperscript{nd} order: $J(J-1)/2$ coefficients}\;.
\end{align*}
Because $J$ is the logarithm of the dynamical range of scales in the field, the number of scattering coefficients increases slowly with the field size. They form a relatively compact set of descriptors. 
In the two-dimension case, one may probe $L$ orientations by using wavelets with angular size $\pi /L$ in Fourier space, whose position angles are $\pi l/L$, with $0 \le l < L$. Compared to 1D cases, there are $L^n$ times more coefficients at the $n$th order.

\subsubsection*{Averaging over orientations}

When considering statistically isotropic fields, there are different levels of orientation reductions. The most extreme one is to average over all orientation indices, which reduces the number of coefficients by an order of $L^n$ and creates a much more compact set of statistical descriptors:
\begin{align}
\label{eq:s1s2}
    s_1(j_1) &\equiv \langle\, S_1(j_1,l_1)\,\rangle _{l_1}\\
    s_{2}(j_1,j_2) &\equiv \langle\, S_2(j_1,l_1,j_2,l_2)\,\rangle _{l_1, l_2}\,,
\end{align}
where $\langle \cdot \rangle_l$ denotes an average over orientation indices. As an example, for an image with a 256 $\times$ 256 pixels size, there are only 29 reduced scattering coefficients ($S_0, s_1, s_{2}$) when the full range of scales ($J$ = 7) is probed.

However, these reduced coefficients $s_n(j_1, ..., j_n)$ only depend on scales and do not provide any `shape' or morphological information information which depend on angles. A less aggressive reduction of the 2nd-order coefficients is to keep the angular dependence on $l_2-l_1$ and average only over $l_1$. This makes the summary statistics invariant to rotation while preserving morphological information. It reduces the number of coefficients by $L$ instead of $L^n$. For example, for an image with 256 pixels on the side, there are 21$\times L$ such coefficients.
Finally, an even more informative reduction can be obtained by applying the scattering idea to angular dependencies, as introduced in \citep{Sifre_2013}.

\subsubsection*{Normalization}

At all orders, the scattering coefficients possess the units of the input field $I_0$.
It is sometimes convenient to manipulate dimensionless statistics. In addition, as the coefficients $S_n$ are proportional to their previous-order field $I_{n-1}$, they are correlated. In order to deal with unitless coefficients, de-correlate them, and ease their interpretation, one can use the normalized scattering coefficients given by
\begin{equation}
s_n = \frac{S_n}{S_{n-1}}\,,
\end{equation}
\citep{Bruna_2015}. This is similar to the normalization commonly found in the context of moment statistics, such as the dimensionless skewness and kurtosis. Note that in the equation above $S_n$ and $S_{n-1}$ should belong to the same branch of the scattering tree, i.e., having the same $\{j_1,...,j_{n-1}\}$. The 1st-order ratio $S_1/S_0$ is not always necessary, depending on how the mean of the field is defined.

\begin{figure*}
    \centering
    \includegraphics[width=\textwidth]{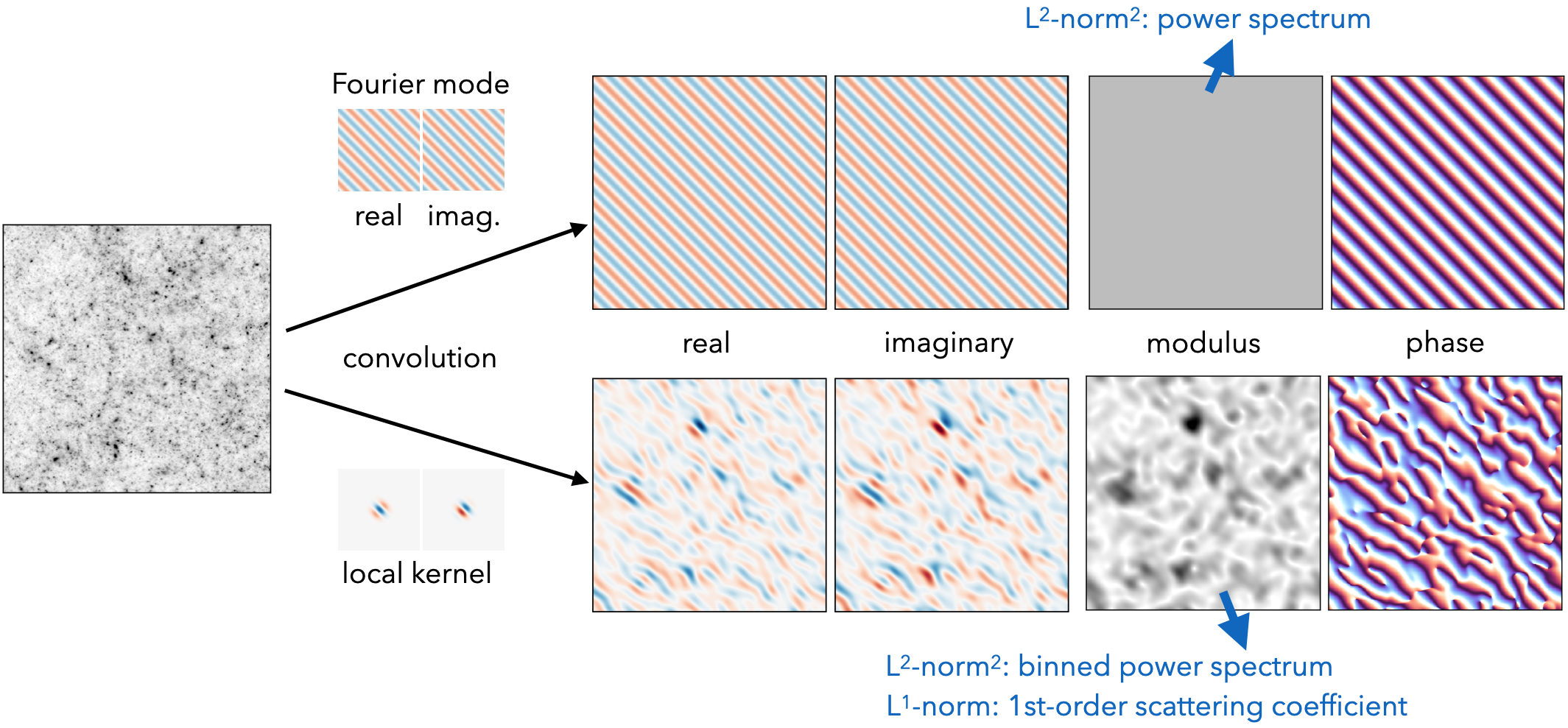}
    \caption{A visual comparison of the estimation of the power spectrum at one 2d frequency and a scattering coefficient at one scale and orientation.}
    % convolutions with Fourier and wavelet kernels and an illustration of the role of modulus. The power spectrum at a single frequency can be calculated by a Fourier convolution plus an L$^2$ norm. Similarly, a 1st-order scattering coefficient $S_1$ is calculated by a wavelet convolution plus an L$^1$ norm.}
    \label{fig:Fourier_vs_wavelet_convolution}
\end{figure*}

\subsection{Key properties}
\label{sec:properties}

The scattering transform generates a statistical description of a field with a number of attractive properties for extracting information from data. It shares all the desirable properties listed for the power spectrum in section~\ref{sec:PS}, namely: translation invariance, rotation invariance (obtained by angular averaging in Fourier space), dimensionality reduction, scale separation, energy extraction. Here, we highlight additional properties of this estimator which offer key advantages for scientific analyses:
\begin{itemize}

\item \textbf{Extraction of morphological information:}
The raison d'\^etre of the scattering transform is its ability to extract non-gaussian information and thus characterize some of the morphology of structures present in a field. It does so by being able to capture interactions between scales, in contrast to the power spectrum. The fast convergence of the energy/information extraction enables one, for a wide range of physical fields, to quantify the relevant morphological information using only a low-order scattering transform. This is illustrated in figures~\ref{fig:PS_limitation} and \ref{fig:synthesis} where second-order scattering transforms are used.

\item \textbf{Another way to describe the energy distribution:}
energy-preserving representations are often important, in particular to physicists. When the so-called admissible wavelets (see Appendix) are used, the energy or variance (L$^2$-norm$^2$) is exactly partitioned into scattering coefficients \citep{Mallat_2012}:
\begin{eqnarray}
\label{eq:energy_partition}
\langle |I_0|^2 \rangle &=& \left\|S \right\|_2^{2}=
S_0^2 + \sum_{j_1} S_1^2 + \sum_{j_1,j_2} S_2^2 + \,\cdots
% \nonumber
\end{eqnarray}
where $S$ is the set of all scattering coefficients of $I_0$\footnote{This partition is an infinite expansion. In practice, one can only work with a truncated expansion but it is interesting to point out that one can decompose the energy into a finite set of low-order scattering coefficients together with the next-order \emph{scattering fields} $I_{n+1}$:
\begin{align}
    \langle |I_0|^2 \rangle = S_0^2 + \sum_{j_1} S_1^2 + ... + \sum_{j_1\cdots j_n} S_n^2 + \sum_{j_1\cdots j_n} \langle |I_{n+1} |^2 \rangle\,.
    % \label{eq:energy_partition}
\end{align}
}
% Note that with $S_0$ excluded, the others add up to the total variance of $I_0$.
At each order of the scattering transform, energy is moved towards lower frequency by the modulus and some of it is extracted by the average over the field. The remaining part is further shifted towards lower frequencies by a new application of the modulus of the wavelet transform. Repeating this process guarantees the extraction of all the energy of the input field. It is also important to point out that the energy of scattering coefficients with order at least $n$ comes almost only from higher frequency ranges.

\item \textbf{Fast convergence and compactness:} An extremely useful property of the energy partition is that, for a wide range of relevant fields, the norm of scattering coefficients decays exponentially fast when their order increases. More precisely, this exponential decay of the coefficients is guaranteed as soon as the Fourier transform of $I$ decays at least as fast as $O(|k|^{-1})$, as proved by Waldspurger et al. (2017). As a result, in many practical applications, \emph{the leading orders of the scattering transform is sufficient to extract relevant information} from a field. One can thus work with a compact yet powerful set of summary statistics. This was shown by Andén \& Mallat with a database of audio signals and \citet{Bruna_2013} with the Caltech-101 image texture database. Numerous applications of the scattering transform indicate that only a sub-percent fraction of the energy remains to be extracted by orders greater than three. In this paper, we therefore focus on only second-order scattering transforms. The visual examples displayed in figure~\ref{fig:synthesis}, which are all computed using only a second-order scattering transform, convincingly show that the relevant perceptual information is sufficiently well captured. A description of the synthesis step is provided in appendix~\ref{app:synthesis}.

\item \textbf{Stability and asymptotic normality:} 
are important properties for both regression and classification problems. A small perturbation of the input field should ideally yield a small change in the descriptors. The scattering transform has proven stability to adding noise and to geometrical deformations, thanks to the `low-order' non-linear operation, modulus, and the logarithmic binning of scales performed by wavelets. For comparison, higher-order statistics are not stable to additive noise, which means they are sensitive to outliers. This property is not met by Fourier coefficients, unless one averages them using logarithmic bins of scales and regular frequency kernels as opposed to sharp bins using $k_{\rm min}, k_{\rm max}$ values. In addition, by using a low-order non-linearity (modulus), the scattering transform never amplifies the tail of the field pdf. The average over the field thus gaussianizes quicker than the average involved in the power spectrum, and much quicker than the higher-order moment-based statistics. This property facilitates likelihood parameterization and improves inference robustness.
\end{itemize}

\subsection{Understanding the scattering operations}

We now provide some insights into the key operations used in the scattering transform. We remind the reader that we are are interested in extracting information from stationary fields (or textures), i.e. fields invariant to translation.

\subsubsection{An overview on translation-invariant descriptors}
\label{sec:translation_invariant}

For simplicity and interpretability, it is desirable to use a descriptor obtained from linear operations, if possible. However, if translation invariance is required, only one such descriptor can be constructed: the global mean of the field. The reason is as follows: let us consider a field $I(x)$ as a vector in function space. Translations are orthogonal linear operations of $I$. They share the same set of eigen vectors: the Fourier modes with eigen values $e^{-ik\Delta x}$. Translation invariance requires an eigen value of unity, which can only be obtained for $k=0$, corresponding to the mean of the field over $x$. Therefore, to go beyond the mean, an estimator must involve a \emph{non-linearity}. To be informative and translation invariant, such a non-linear operation typically moves high-frequency power into the $k=0$ mode. This non-linear transform must not interfere with the required `translation invariant' property, in the sense that it should \emph{commute} with translations (called `equivariant'). A simple strategy is to use a pointwise non-linear operation.

In summary, there is an overall strategy to extract translation-invariant information beyond the trivial mean of a field:
\begin{itemize}
    \item One can, if desired, focus on specific scales of features by using equivariant linear operations: the convolution, which is naturally found in the power spectrum, scattering transform, CNNs, etc.
    \item One must use equivariant non-linear operations, such as a pointwise modulus, squared modulus, or ReLU, etc, to go beyond the mean. This operation can be unary (taking one argument as input), such as the modulus used in the scattering transform and the activation functions in CNNs. It can also be binary or $n-$nary, such as the multiplications in moment-based statistics. 
    \item One can then take the mean (or max) over the field to extract a translation-invariant quantity.
\end{itemize}

\subsubsection{Why using complex wavelets?}

The scattering transform uses complex-valued (also called analytical) wavelets as convolution kernels. They are localised waveforms and band-pass filters with similar geometric shapes but different sizes or orientations. They can be used to select fluctuations around certain frequencies, as a function of position. By convolving a field with a family of wavelets, it provides a linear transform which expresses the presence certain frequencies as a function of position. It is best suited for fields composed of a sparse superposition of localized structures at different scales, such as peaks, edges, patches, eddies, filaments, etc, or transient features in time series. These localized features create interaction across neighbouring frequencies. They can also be oriented. A wavelet representation usually reveals them more clearly than in the original pixel space or in Fourier space.

It is interesting to point out that such a family of dilated and rotated wavelets naturally emerge in `optimized' image-coding schemes: it has been known since \citet{Hubel_1968} that oriented wavelet-like kernels are found in the receptive fields of the visual cortex of animals. They also naturally emerge when a learning algorithm attempts to find a sparse linear code for natural scenes \citep{Olshausen_1996}. Similarly, they are found in the first layer of CNNs \citep{Krizhevsky_2012}. This universality strongly suggests the use of wavelets in the design of a generic estimator aimed at characterizing a wide range of physical fields. 
% [?some words about 1D/audio]

The kernels found in the mammalian visual cortex, sparse linear codes and CNNs are real-valued quantities as opposed to the complex-valued wavelets used in the scattering transform. Interestingly, for these three image-coding schemes, each type of kernel is found in both its symmetric and asymmetric versions. The use of complex-valued wavelets therefore appears to be an equivalent description for which the symmetry or, similarly, the local displacement is simply carried by the complex phase. Both systems have the same capacity to describe patterns.

Finally, as already mentioned, the dilated wavelets used in the scattering transform allow for a logarithmic sampling of scales or, in other words, a logarithmic `tiling' of the Fourier space. This significantly reduces the number of scattering coefficients to be calculated and provides the foundation of deformation stability.

\subsubsection{The role of the modulus}

The complex modulus allows to estimate the presence or strength of a wavelet-like feature while discarding information on its precise position (which is carried by the complex phase). As shown in figure~\ref{fig:Fourier_vs_wavelet_convolution}, the convolution of the input field $I(x)$ by a complex wavelet $\psi(x)$ localized in real space and spanning a factor two in frequencies around $k_0$ provides us with real and imaginary values of the strength of symmetric and anti-symmetric oscillations in $I$ around that frequency at each position $x$. The peaks of these two estimates are offset by $\delta x\approx\pi/2k_0$. By combining them into a single real-valued field, the modulus forms an estimate of the \emph{local strengths} of fluctuations with frequency around $k_0$, irrespective of their actual centroid within a region of size $\delta x\simeq \pi/k_0$, so in general the field after modulus $|I\star\psi|$ has lower frequencies than the original fluctuations in both real and imaginary parts, In other words, the modulus $I\star\psi \rightarrow |I\star\psi|$ `scatters' fluctuations, information, and energy from high-frequency into lower-frequency and zero-frequency (translation invariant) modes.

It has several implications. The first is stability to deformation. To understand it, let us considering small deformations, i.e. deformation that can be approximated by a local shift $|\delta x|<\pi/k_0$. When such a deformation occurs, while the phase of the complex quantity $I \star \psi(x)$ may vary by order unity, the modulus is stable.
% when a small perturbation $|\delta x|<\pi/k_0$ is introduced, 
This shows that \emph{locally} discarding phase information is similar to discarding information on the exact position of fluctuations within the envelope of the wavelet, of size $\pi/k_0$. This provides one of the important stability properties of the scattering transform. The role played by the modulus in the context of image classification has been studied by \citet{Guth_2021}.
% Although a wavelet modulus operator removes the complex phase, it does not lose information because the variation of the multi-scale envelopes is kept.
% \footnote{BM: talk about modulus concentrating frequencies by a factor 2, leading to an exponential decay.}
%
% What does a `low frequency' field mean? It basically means more stability to deformation, which is a key property structures. When we look at the dimensionality of a field, we realize that most of the degrees of freedom are located at high frequencies (each Fourier mode is a degree of freedom).

Another implication is the sensitivity to sparsity. The sparser the field, the more scattering operations are needed in order to move the energy or information towards $k=0$ where it is extracted by the global mean. Therefore, the ratio between scattering coefficients at different orders provides us with a sparsity estimate of the field.

Finally, as the modulus preserves the norm of the field, the scattering transform creates an energy partition of the field. This is desired both in terms of physical interpretations and for stability and robustness.

\subsubsection{Scale interactions}

As mentioned in section~\ref{sec:P_to_S}, the first-order scattering coefficients are qualitatively similar to amplitudes of a binned power spectrum $P(k) = \langle |I \star e^{-ikx}|^2 \rangle$.
The second-order coefficients extract more information about the field $I$ by re-applying the scattering operation.
Thus, the second order scattering is similar to a power-spectrum analysis of the locally-measured power spectrum field. This approach creates a hierarchy of analysis and coefficients. It can be understood as using a hierarchical assembly of simple structures to characterise more complicated ones.

When talking about statistics beyond the power spectrum, we often describe them as `probing scale interactions', as opposed to the power spectrum $P(k)$ which treats scales separately as Fourier modes. It is important to point out that the scattering transform is only sensitive to a subset of all possible scale interactions. From figure~\ref{fig:ST_tree} we notice that the scattering is performed along a tree structure without `cross-talk' between different branches. By construction, the second-order coefficients $S_2(j_1, j_2)$ cannot probe interactions between pairs of arbitrary Fourier modes. They can only capture interactions of Fourier modes within the pass-band of the first wavelet $\psi^{j_1}$ (see Appendix~\ref{app:S_to_Npoint} for an illustration). This limits its ability to characterize fields with sharp features. Indeed, the sharper the feature, the wider its spread in Fourier space and the less it can be captured within one wavelet pass-band. Despite this limitation, the scattering transform is powerful enough to capture the relevant morphological information in a wide family of scientifically-relevant fields, as illustrated in figure~\ref{fig:synthesis}.

\begin{figure}
    \centering
    \includegraphics[width=\columnwidth]{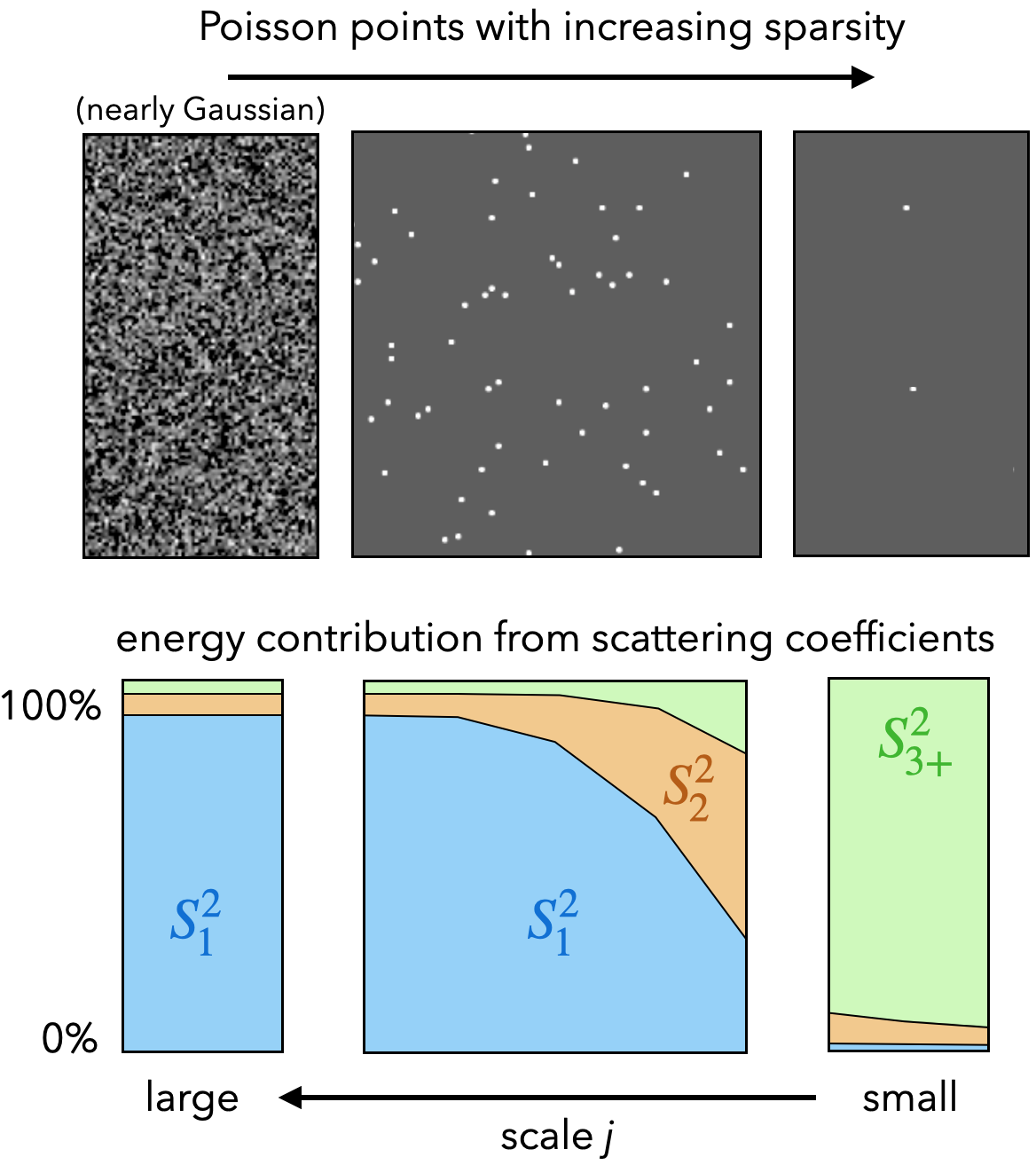}
    \caption{Poisson processes with varying density and the corresponding variance (energy) partition. 
    As the field becomes more sparse, the energy described by the scattering coefficients moves to higher order.
    In the intermediate regime, which is relevant to most fields in scientific research, a substantial fraction of the energy is carried by the first- and second-order coefficients.}
    \label{fig:energy}
\end{figure}

\section{Interpretability}
\label{sec:interpretability}

Being mathematically well-defined, the scattering coefficients also have interesting interpretations which are related to scale, energy, feature sparsity and shapes. A scattering transform expansion up to a finite order does not uniquely determine a signal but characterizes some of its statistical properties. It is impossible to exactly recover the input field by only the scattering coefficients. However, a perceptually plausible sample of the same texture can often be generated, as illustrated in figure~\ref{fig:synthesis}.

\subsection{First-order coefficients $S_1$}
\label{sec:interpretability_S1}

The $S_1$ coefficients are qualitatively similar to the power spectrum amplitudes. Both of them characterize fluctuation strength as a function of scale. They differ only in two aspects: the scattering transform selects fluctuations using a family of dilated overlapping wavelets rather than potentially arbitrary and narrow bins in scale. Second, while the power spectrum uses the L$^2$ norm of the convolved field, the scattering transform uses the L$^1$ norm which does not amplify fluctuations and contributes to minimizing the variance of the estimator.

To understand how to interpret the coefficients, we first apply the scattering transform to a series of random point processes with densities ranging from sparse to dense. By construction, they all have a flat power spectrum which, as the point density increases, asymptotically corresponds to Gaussian white noise. We point out that the majority of physical fields, after a convolution by a localized kernel representing the typical features, would lie in the intermediate regime. The amplitudes of the corresponding scattering coefficients are shown in figure~\ref{fig:energy}. The upper panels illustrate realizations of these fields and the bottom panels show how the energy is partitioned through the first, second, and higher-order scattering coefficients. As expected, in the case of a Gaussian random field, most of the energy is carried by the first-order coefficients. When the density of the Poisson process decreases, the sparsity of the field increases. The partition of energy now spreads towards second- and higher-order scattering coefficients. This effect is scale-dependent and more pronounced on small scales. Indeed, considering large scales is similar to smoothing the field on those scales, thus reducing the level of shot noise and making the fluctuations closer to Gaussian. Interestingly, we observe that, in this intermediate regime, the fraction of energy carried by scattering coefficients at orders higher than two is relatively small. For an extremely sparse field, as shown in the right panel, only a negligible fraction of the energy is stored in low-order coefficients. However, for such a field, the statistical representation becomes less necessary.

\subsection{Second-order coefficients $S_2$}

Second-order scattering coefficients offer deeper insight into some of the statistical properties of a field. They characterize transient phenomena such as localized structures in space or time, such as amplitude modulations or rapid changes. Substantial non-Gaussian information is carried by these coefficients. In particular, they provide co-occurrence information at the scales $j_1$ and $j_2$ and thus capture interferences of the field between features selected with two successive wavelets $\psi_{j_1}$ and $\psi_{j_2}$. This is why they are called scattering coefficients.

The second-order coefficients are obtained after applying the scattering operation to the transformed field $I_1$. They characterize the assembly or clustering of \emph{patterns} (with a given scale $j_1$) by quantifying the strength of their fluctuations mapped in $I_1^{j_1}$ as a function of scale $j_2>j_1$. In other words, this corresponds to the clustering, on a scale $j_2$, of structures on scale $j_1$. Being a function of the two scales $j_1$ and $j_2$ and two orientations $l_1$ and $l_2$, the second-order coefficients are more numerous than at first-order. However, as they are significantly correlated, dimensionality reduction techniques can be used to construct a more compact set of summary statistics, if need be.

\begin{figure}
    \centering
    \includegraphics[width=\columnwidth]{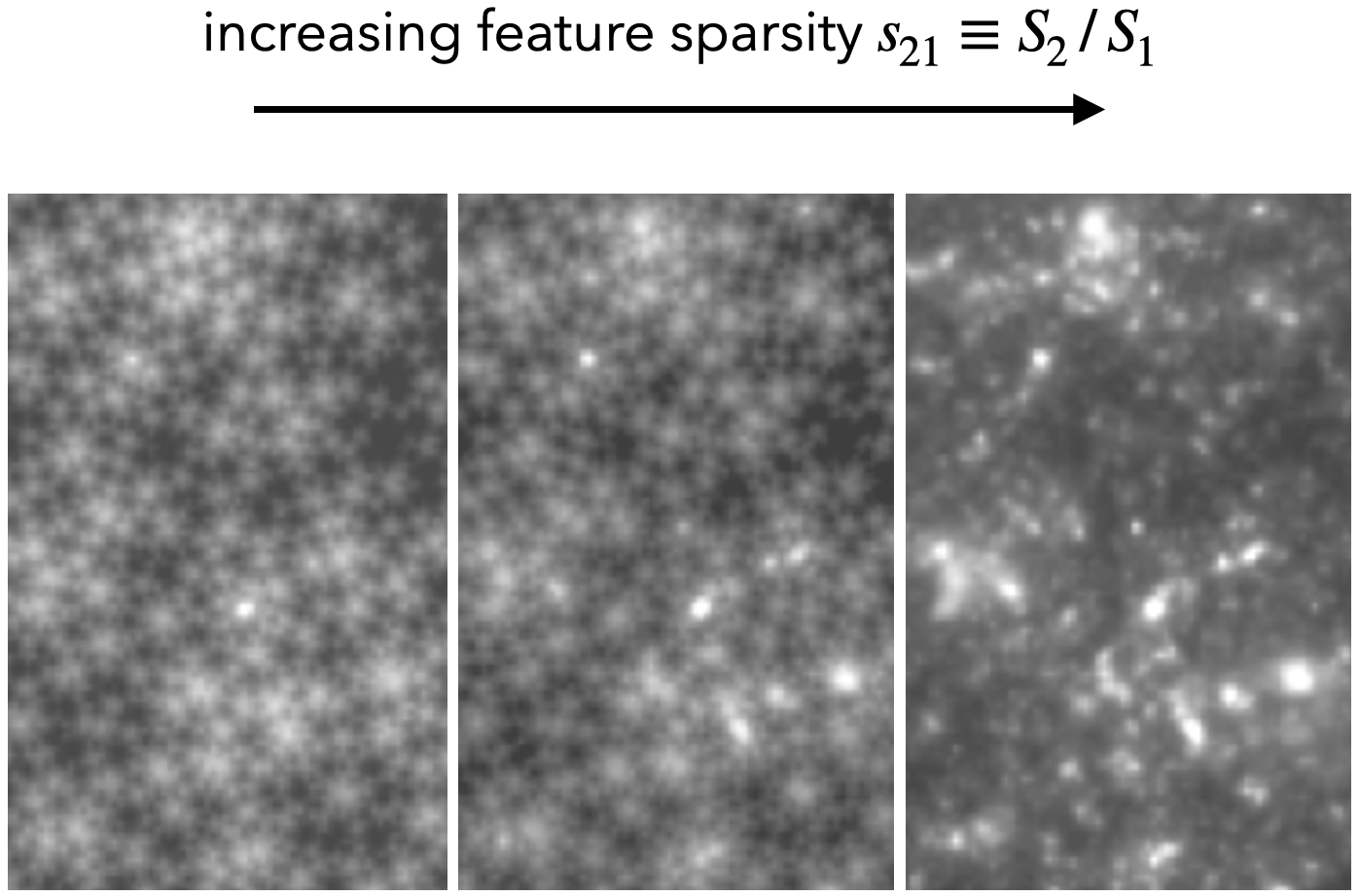}
    \caption{Syntheses with different values of the reduced scattering coefficients $s_{21}$ illustrating their connection to field sparsity.}
    % A high $s_{21}$ means that the structures are sparsely distributed in the field, whereas a low value corresponds to a Gaussian-like texture, where fluctuations are widely spread. Note that the point-like feature shown here is only one example of sparse features. Other examples are shown in figure~\ref{fig:s22}.}
    \label{fig:s21_new}
\end{figure}

\begin{figure}
    \centering
    \includegraphics[width=\columnwidth]{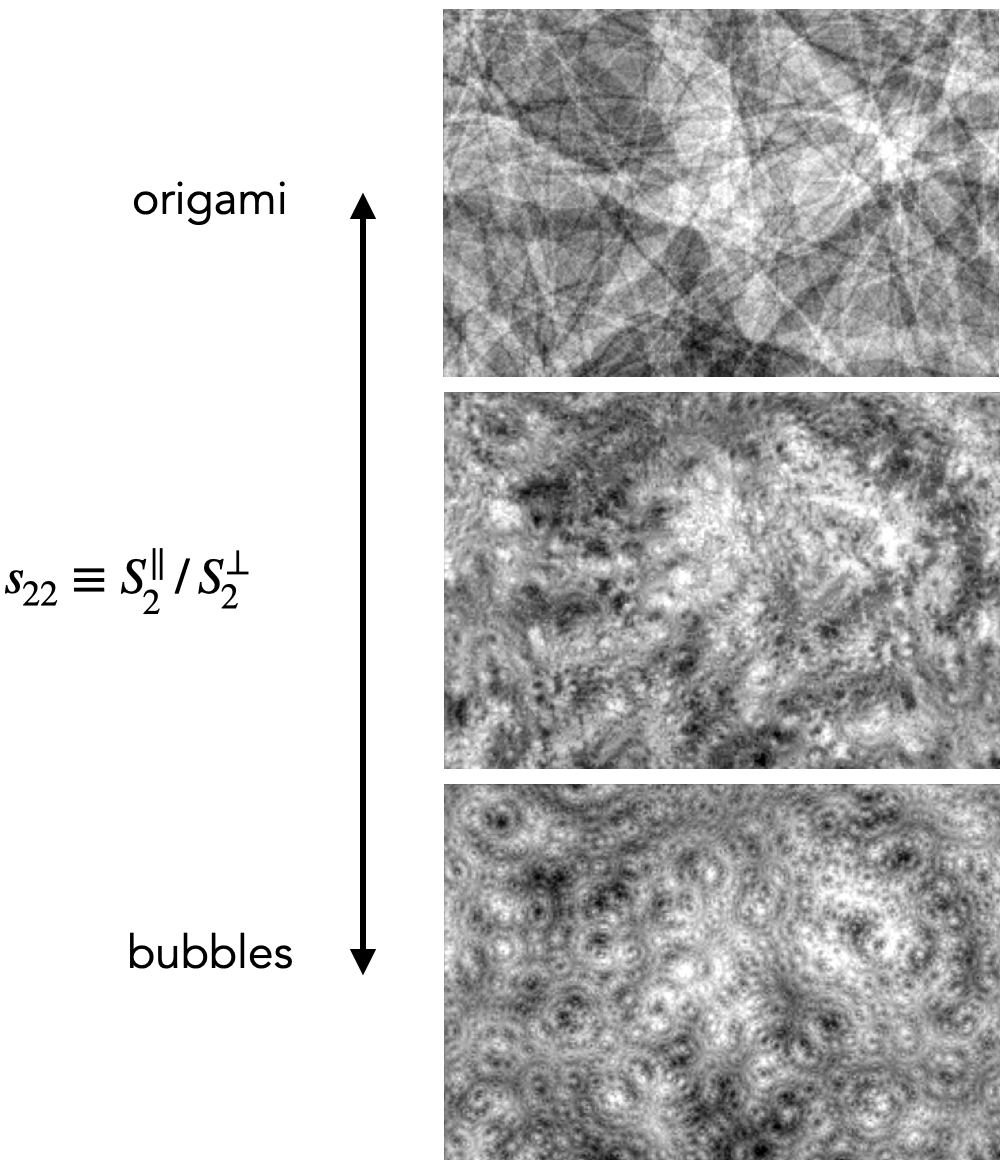}
    \caption{Syntheses with different values of the reduced scattering coefficients $s_{22}$ illustrating their connection to shape properties. Varying $s_{22}$ from about $0.5$ to $2$ changes the shapes of the dominant features in the generated textures from swirls and bubbles to wrinkles and filaments.}
    \label{fig:s22}
\end{figure}

In order to interpret some of the second-order scattering coefficients, one can focus on two particularly meaningful combinations:
\begin{itemize}
    \item \textbf{Feature sparsity $s_{21}$:} the ratio between $S_2$ and $S_1$ coefficients can be interpreted as an estimate of feature \emph{sparsity} as a function of scales, as illustrated in figure~\ref{fig:s21_new}. Intuitively, sparsity indicates whether fluctuations or structures are concentrated at a few positions or widely spread. If no relevant orientation information is expected, it is useful to consider the orientation-averaged ratio
    \begin{equation}
        s_{21} \equiv \langle S_2\,/\,S_1 \rangle _{l_1, l_2}\;.
    \end{equation}
    This quantity directly informs us on departures from Gaussian random fields. It is sensitive to the strength of structure or localized features present in the field. If the density of such features were to increase and start to overlap to the point of becoming ambiguous, the central limit theorem will apply and the field will Gaussianize. During this process $s_{21}$ will decrease and approach unity.
    Sparsity and structures are related to lower entropy, compared to the level reachable with the same energy but through a Gaussian random field. Unlike the estimation of $L^2$-norm or energy, sparsity estimates are not generic and depend on the chosen basis. Using wavelets allows one to approach the problem generically. Maximizing the sensitivity to a specific manifestation of sparsity would require optimizing (or learning) specific filter shapes.

    \item \textbf{Shape $s_{22}$:}
    the shape of features or fluctuations is often an important aspect of a field. This is usually the property that allows the identification of textures around us, despite changes in angle, illumination, scale, etc. Shape information is captured by a number of the $S_2(j_1,l_1, j_2,l_2)$ coefficients but it is possible to introduce a convenient reduction which carries valuable information. If rotation invariance applies, the global orientation can be averaged out but the relative angle between $l_1$ and $l_2$ remains informative. We can probe such a dependence by selecting two key orientations -- parallel and perpendicular -- and consider the ratios of their respective scale-dependent coefficients as a new group of reduced coefficients:
    \begin{equation}
        s_{22}\equiv \langle S_2^{l_1=l_2} / S_2^{l_1 \perp l_2} \rangle _{l_1}\;.
    \end{equation}
    This quantity is easily interpretable, as illustrated in figure~\ref{fig:s22}. Fields with $s_{22} < 1$ display curvy, soft-look patterns like bubbles and swirls, as the anisotropic fluctuations at one scale ($j_1$) are mainly distributed along the orthogonal direction with respect to larger scales ($j_2$).
    Fields with $s_{22} > 1$ display straight, hard-looking patterns like long lines, filaments, and origami textures. For those, the anisotropic fluctuations are mainly distributed along the same direction. The visualisation in figure~\ref{fig:s22} is achieved by randomly generating images with the same $S_1$ and $s_{21}$, i.e., similar power spectrum and feature sparsity, but different $s_{22}$ coefficients. 
    Technically, we set the scale dependence of $S_1$ and $s_{21}$ as power laws ($\propto 2^{a{j_1}+b{j_2}}$). In each panel, we set no scale dependence for $s_{22}$ to acquire the basic intuition. In real world, these $s_{22}$ coefficients can of course depend on the scale combinations ($j_1,j_2$), resulting in rich possibilities of textures. Note that similar to $s_{21}$ (a measure of sparsity), $s_{22}$ also depends only on the two scales $j_1$ and $j_2$. So, after including $s_{22}$, the set of reduced scattering statistics becomes even more powerful while remaining compact.
\end{itemize}

\begin{figure*}
    \includegraphics[width=\textwidth]{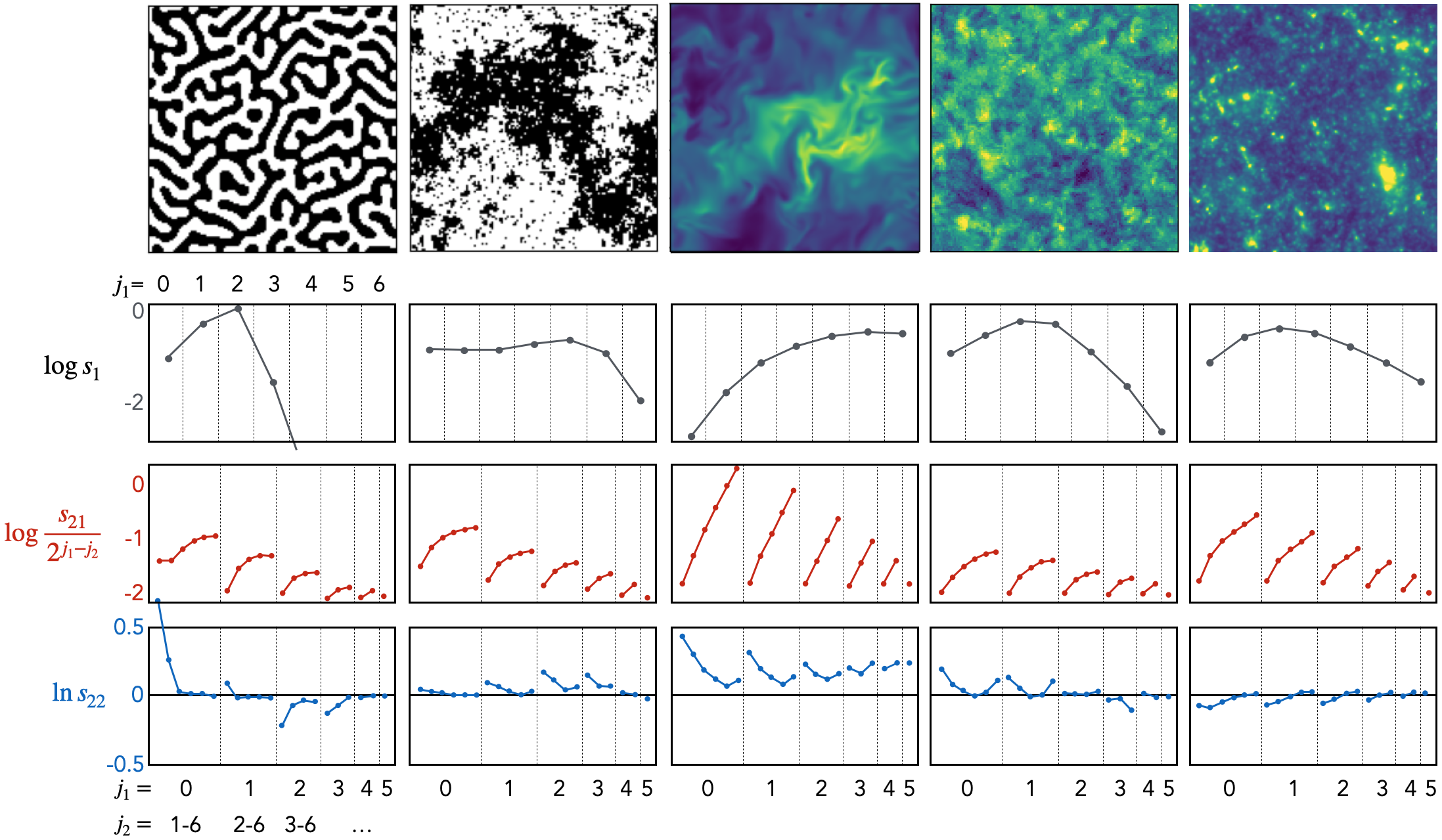}
    \caption{
    The scattering coefficients of various fields or textures, as a function of scale. Intuitively, the $s_1$ coefficients characterize the amplitudes of fluctuations, the $s_{21}$ coefficients characterize the spatial sparsity of structures averaged over orientations, and the $s_{22}$ ones characterize the shape of structures (curvy vs. straight). The $s_{21}$ coefficients have been normalized by $2^{-j_2}$ which is the asymptotic behaviour of many ergodic fields. Note that we show the mean value of the coefficients averaged over 50 realizations of each process to reduce uninformative fluctuations.
    %We also use base-2 logarithmic scale for the $s_1$ and $s_{21}$ coefficients.
    }
    \label{fig:examples}
\end{figure*}

\section{Relation to other methods}

\subsection{Comparison to higher-order statistics}
\label{sec:higher-order}

One way to extract non-gaussian information is to use higher-order statistics. These higher-order statistics such as $N$-point functions, moments, and poly-spectra are successive multiplications of field variables $\langle I(x_2)I(x_2)..I(x_n) \rangle$ and their linear combinations. They naturally emerge from perturbation theory which has been a motivation to use them, for example in cosmology.

Unfortunately, their practical use is plagued with a number of problems: i) the successive multiplications amplify the variability of the field, leading to a non-robust statistic with rapidly increasing variance and a high sensitivity to potential outliers. This is a severe limitation in the study of 'complex' systems, for which heavy-tailed distributions are ubiquitous; ii) considering $N$-point correlation functions or polyspectra, the number of statistics explodes with $N$, typically diluting the information across a large number of coefficients.

The scattering transform has the ability to circumvent some of these limitations by not raising the power index of the input field and by using the modulus as the required non-linearity to extract information beyond the mean, in contrast to the successive multiplications used by higher-order statistics. The modulus guarantees stability and convergence. We explain this behavior in more detail in appendix~\ref{app:folding}. Another difference originates from the dilated wavelets used by the scattering transform leading to a logarithmically-spaced tiling of Fourier space. This strategy significantly limits the number of extracted coefficients.

Finally, we point out that moment expansions seem to lure some in hoping that information about a probability distribution function can be extracted through its series of moments with arbitrarily high precision. Unfortunately, this approach can fail, even if all higher-order moments are taken into account \citep{Carron_2011}. This problematic regime is met when the distribution is heavy tailed, which is the case of a wide range of fields.

\subsection{Comparison to convolutional neural networks}
\label{sec:cnn}

The scattering transform and CNNs share a number of properties: both of them use localized convolution kernels, non-expansive non-linearity and a hierarchical structure. CNNs are typically decomposed as having a set of $N$ layers with learnable filters aimed at approximating an arbitrary function or, in other words, at building a non-linear representation with a set of coefficients, together with a 'last', fully connected layer which performs a mapping from these features to the final outputs. The coefficients describing both parts are optimized simultaneously during a training phase. The scattering transform, on the other hand, uses preset wavelets as convolutional kernels and is typically used at low order (mainly second order in our case). It can be viewed as a non-trainable shallow CNN. In the scattering transform's approach, the last fully-connected layer is supplanted by traditional regression techniques. 

We point out that while both the number of layers in a CNN and the order of the scattering transform set the number of non-linearities involved in their internal description, the two approaches treat scales differently. In a CNN, the accessible scale increases with depth as the receptive field grows together with decreasing resolution. In contrast, the scattering transform treats a range of scales at each order. The order of the scattering transform is therefore not direclty comparable to the number of layers in a CNN.

The training of a CNN often comes at a substantial computational cost and involves the tuning of a number of hyperparameters. This process is not always guided by well understood principles. Ultimately, the reward can be high: the properly optimized kernels often lead to state-of-the-art performance for tasks involving complex data such as classifying different types of rabbits. Unfortunately, in some cases the over-parametrization of CNNs can lead to a much more brittle statistical model \citep{Szegedy_2013, bruna2019multiscale}. Over-parametrized models tend to over-fit, i.e., to `remember' single realizations instead of generalizing the overall property of the whole training set. Thus the over-parametrized CNNs require a large number of simulations as training set to alleviate the over-fitting problem. With its pre-set wavelet kernels, the scattering transform is not subject to this problem. It is therefore better suited to extract statistical information from a limited set of inputs. This is a regime often found in scientific applications where large datasets or simulations are too costly.

While the scattering transform uses complex wavelet filters, followed by a modulus and an average, CNNs instead use real-valued filters, typically followed by a Rectified Linear Unit (ReLU) and a max-pooling operation. It is interesting to point out that these two seemingly different choices lead to similar operations. 
As shown by Waldspurger (2015), the two can be related mathematically:
\begin{equation}
\begin{aligned}
   \max\operatorname{pool}(&\operatorname{ReLU}(I \star \operatorname{Re}(\psi))) \approx |I \star \psi(x)|\;.
\end{aligned}
\end{equation}
This equivalence between the scattering transform and a CNN architecture can be understood as follows: if the spatial extend of the max-pooling is larger than $\pi/k$, $\operatorname{Re}(\psi)$ will reach $\pm1$ and applying $\operatorname{ReLU}$ followed by $\max\operatorname{pool}$ will lead to unity. The set of operations found in a CNN thus effectively removes the local phase and extracts the modulus, as done with the scattering transform.

\begin{figure*}
    \centering
    \includegraphics[width=\textwidth]{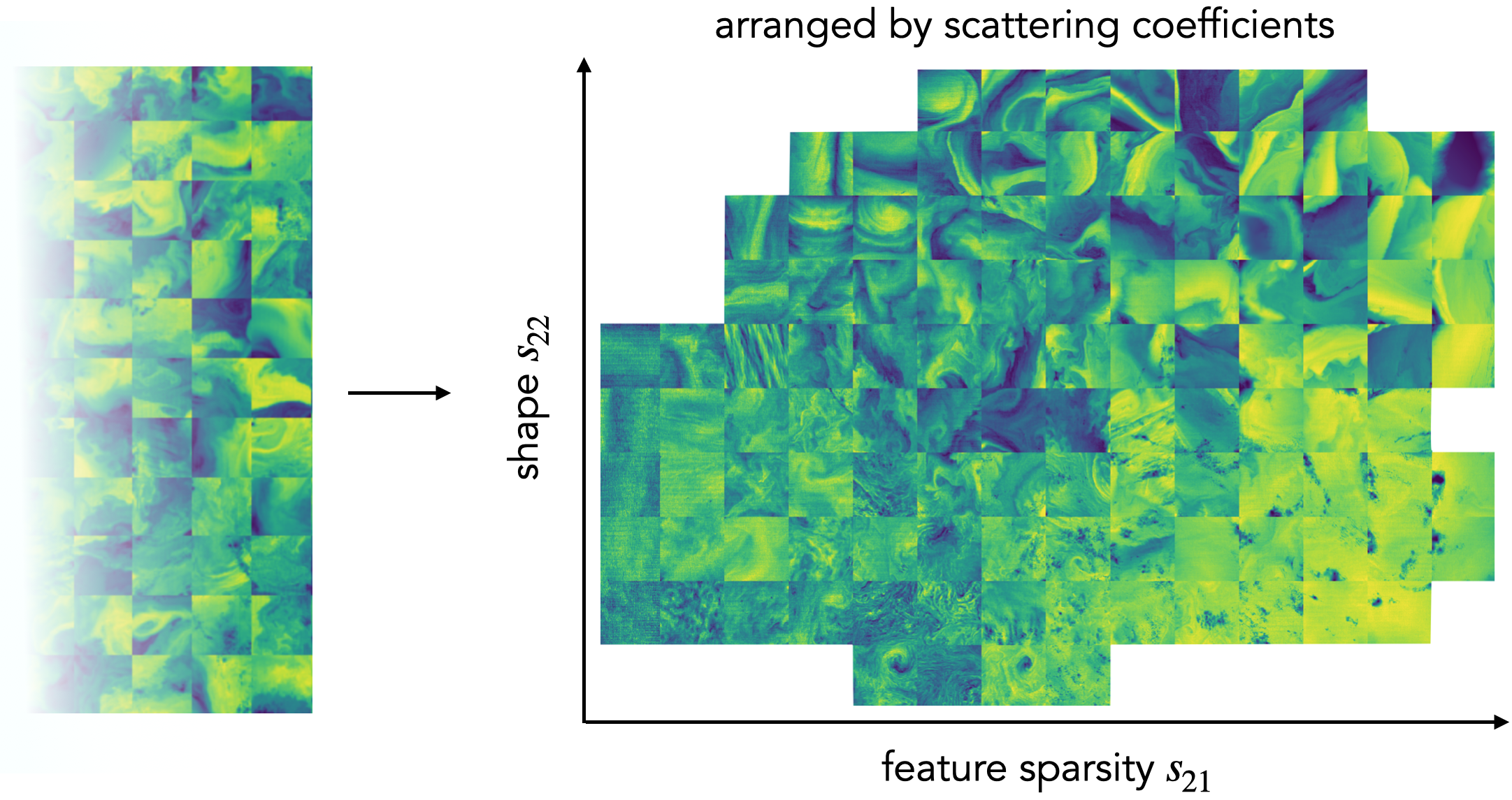}
    \caption{Randomly selected sea temperature fields (left) and a set automatically arranged according to their 2nd-order scattering coefficients (right). The x-axis is the feature sparsity $s_{21}$ averaged over all scales $j_1$ and $j_2$, and the y-axis is that of the shape indicator $s_{22}$.}
    \label{fig:SST}
\end{figure*}

\subsection{Extensions of the scattering transform approach}

\subsubsection*{The phase harmonic estimator and cross-correlations}

The phase harmonic statistic recently introduced by \citet{Mallat_2019} and \citet{Zhang_2021} takes the idea of the scattering transform to the next level. The scattering transform can only probe the scale interactions of a field within a given wavelet in Fourier space. It cannot capture relations between different wavelet footprints using standard cross-correlations, whose expectation values are always zero due to the rapidly fluctuating phases. To address this limitation, one can re-scale the phase fluctuations between two frequencies to match their rate of change and keep their relative coherence. By doing so, cross-correlations no longer have an expectation value of zero. This way to `cross-correlate' fluctuations in different frequency bands can be used to extract scale interactions across different wavelet windows in Fourier space. This approach can also be used to measure non-Gaussian cross-correlations between two different fields. The phase harmonic statistic does provide a richer description of a field and is able to capture scale interactions and morphological information that are not accessible by the scattering transform. 
It has been successfully applied to simulated data in astrophysics \citep{Blancard_2021} and cosmology \citep{Allys_2020} and a variety of point processes \citep{Brochard_2020}.
However, this comes at a cost: the typical number of phase harmonic coefficients is much greater than the number of scattering coefficients, potentially by one or two orders of magnitude. This difference makes data exploration and parameter estimation more challenging with the phase harmonic approach. Finally, it is also important to point out that the scattering coefficients are `first-order' statistics, whereas the phase harmonics are second-order and are therefore subject to higher variance.
% \footnote{we need to cite Tanguy, etc.}

\subsubsection*{Adding trainable modules}

As the complexity of the data increases, a second-order scattering transform might not be sufficient to capture enough information. Considering the scattering transform at higher-order does provide a representation with more accuracy but, as the number of coefficients extracted increases exponentially with the order considered, this becomes quickly impractical. For a given task, one typically does not expect all scattering coefficients to be relevant and, as described in the previous paragraph, one does expect some correlations between coefficients to carry valuable information. In this case, it is possible to use a trainable projector to select informative combinations of coefficients \citep{Zarka_2019}. This can be used to reduce the number of coefficients to work with. Interestingly, by doing so, one obtains the equivalent of a CNN with fixed spatial convolution filters (wavelets) but trainable matrix applied across channels within each layer. This renders the scattering transform much closer to CNN. It can also be used to better understand the structure and behavior of CNNs given that the training phase is restricted to learning a set of correlations but does not involve the optimization of any filter. With such a learning ability, it has been shown that scattering networks can reach the performance of a ResNet-50 on the ImageNet classification task \citep{Zarka_2020, Guth_2021}.

\begin{figure}
    \centering
    \includegraphics[width=0.8\columnwidth]{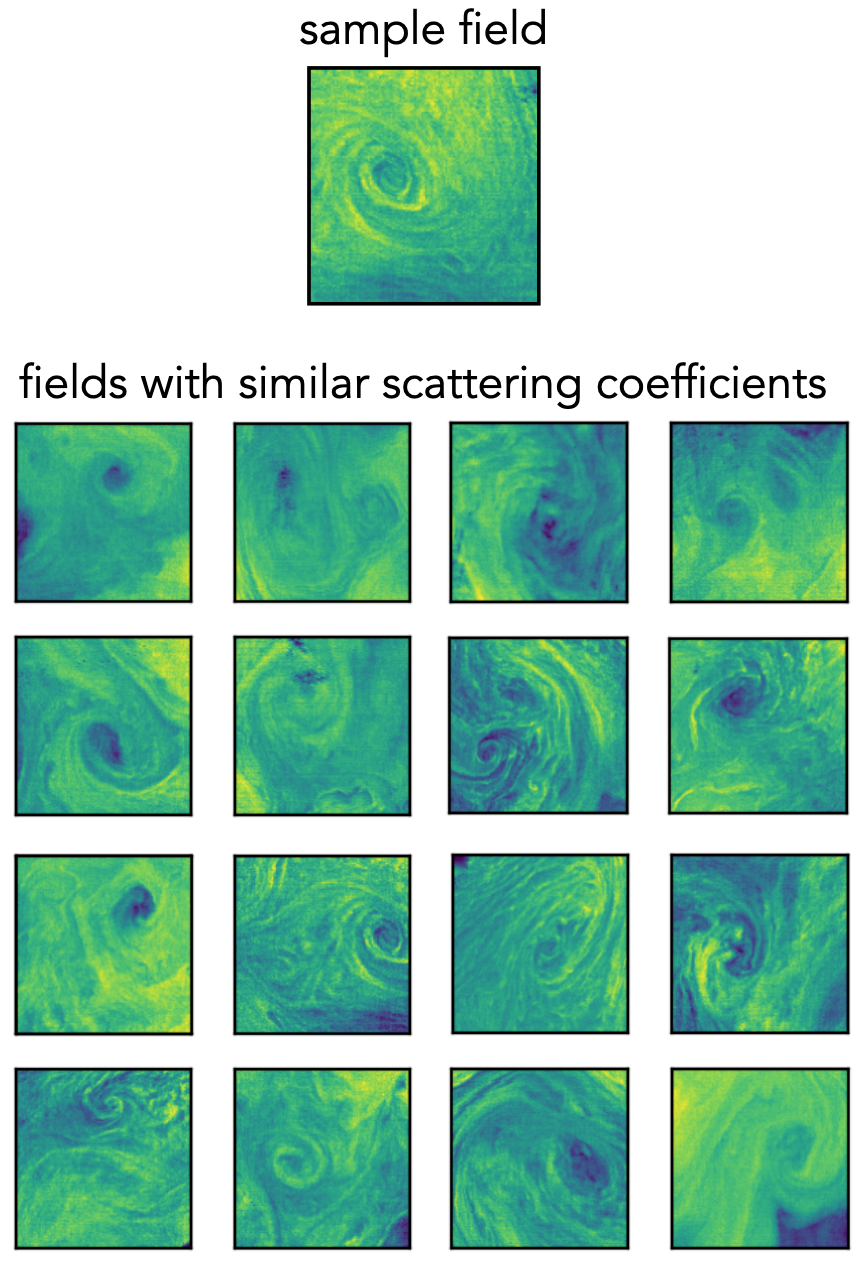}
    \caption{An examples showing a selected sea temperature field with an eddy (top) and a gallery of similar fields selected from their similar $s_{22}$ coefficients -- the shape indicator.}
    \label{fig:SST_similar}
\end{figure}

\section{Applications}

\subsection{Software implementation}

Over the years, a number of software packages have been developed to implement the scattering transform. Recently, the developers of several of those teamed up to create the ``Kymatio'' package \citep{Kymatio_2018}: \href{https://www.kymat.io}{https://www.kymat.io}. It can take advantage of GPU acceleration, provides a variety of different frontends and backends, and has been widely used in the signal processing literature.

In implementations of the scattering transform, a fast Fourier transform (FFT) is used to perform a convolution so that the computation complexity is only $O(N\log N)$ with the number of pixel $N$. The number of FFTs required is roughly equal to that of the scattering coefficients (around $J^2L^2$ to the 2nd order). For physical fields where only a global average of the scattering coefficients are needed, the authors provide a modified version optimized in speed: \href{https://github.com/SihaoCheng/scattering_transform}{https://github.com/SihaoCheng/scattering\_transform}. 
% \footnote{In one sentence, say what is different so the reader has an idea of the origin of the improvement.}
Using a GPU and considering a batch of images with 256 pixels on the side, our code computes scattering coefficients (using all scales and 4 orientations) for about $1,000$ images in one second. This is 4 times faster than using the Kymatio's pytorch backend. On a CPU, it processes about 30 images per second and is 10 times faster than Kymatio.

% with $J=7$ (all dyadic scales) and $L=4$ orientations
% the speed for a batch of images with 256 $\times$ 256 pixels, $J$ = 7, and $L$ = 4, is typically 1000 images per second with our code and 250 images with 

\subsection{Data analysis examples}

Given its close connections to CNNs, the scattering transform has often been used in classification problems. Its scope and usefulness are, however, much broader. It can be used in a wide range of signal processing applications, from exploratory data analysis to precise parameter inference when a model is available  -- two types of situations commonly met by scientists.

\subsubsection*{Exploring fields}

Exploratory data analysis is a task often encountered in scientific research when models and training sets are not  available. In some cases, it even takes places without specific or well-defined questions in mind. When datasets cannot be inspected by eye, a required first step is to organize its elements. As we will show below with two examples, the scattering transform has the ability to do so in a way that is generic, unsupervised and interpretable.

To illustrate such an exploratory process, we use a publicly-available NASA dataset with measurements of the sea surface temperature around the world\footnote{\url{https://www.ghrsst.org/ghrsst-data-services/products/}}. Being a tracer of the water dynamics, the sea surface temperature field exhibits a wide range of complex morphologies. Following \citet{Prochaska_2021}, we will attempt to organize the data and identify interesting rare objects but, in contrast to these authors, we will use the scattering transform instead of a trained neural network. The dataset is a collection of around 100,000 images with 128 pixels on the side, representing a physical distance of around 128 km. The nature of these fields varies wildly: they can be quiet, display interfaces between cold and warm currents, eddies,  etc. as well as small clouds present in the images. The type of a field does not depend on the position of relevant features. It is translation invariant. A technique like principal component analysis is therefore not appropriate to distinguish them. Indeed, the leading variation in the data will be dominated by translations. The scattering transform, however, offers a simple way to meaningfully explore the dataset. To do so, we compute for each image:
\begin{itemize}
    \item a feature sparsity index: $s_{21}\equiv S_2 / S_1$ averaged over all scale combinations $(j_1, j_2)$;
    \item a shape index: $s_{22}\equiv S_2^{l_1 \parallel l_2} / S_2^{l_1 \perp l_2}$ averaged over all scale combinations $(j_1, j_2)$.
\end{itemize}
This representation provides us with a two-dimensional characterization of the statistical properties of each image. We visualize a small subset of the entire distribution in figure~\ref{fig:SST}. As can be seen, our representation meaningfully organizes the different types of fields in a continuous manner. 

This scattering representation can also be used to measure the similarity between fields for data exploration purposes. To illustrate this point, we select one field with a large eddy as shown in figure~\ref{fig:SST_similar}. We then search for the 16 nearest neighbors in our two-dimensional space of scattering coefficients. The corresponding fields are displayed in the figure. Their morphological similarities to the original field are clear. It is interesting to point out that, despite being very rare in the overall dataset, we can easily identify them through their characteristic $s_{22}$ coefficients. In contrast, the \citet{Prochaska_2021} outlier analysis using a CNN combined with a normalizing flow did not manage to identify such fields. In summary, for the proposed task, the scattering transform can outperform a trained neural network. It provides us with physically interpretable results without any training nor supervision.

\begin{figure*}
    \centering
    \includegraphics[width=\textwidth]{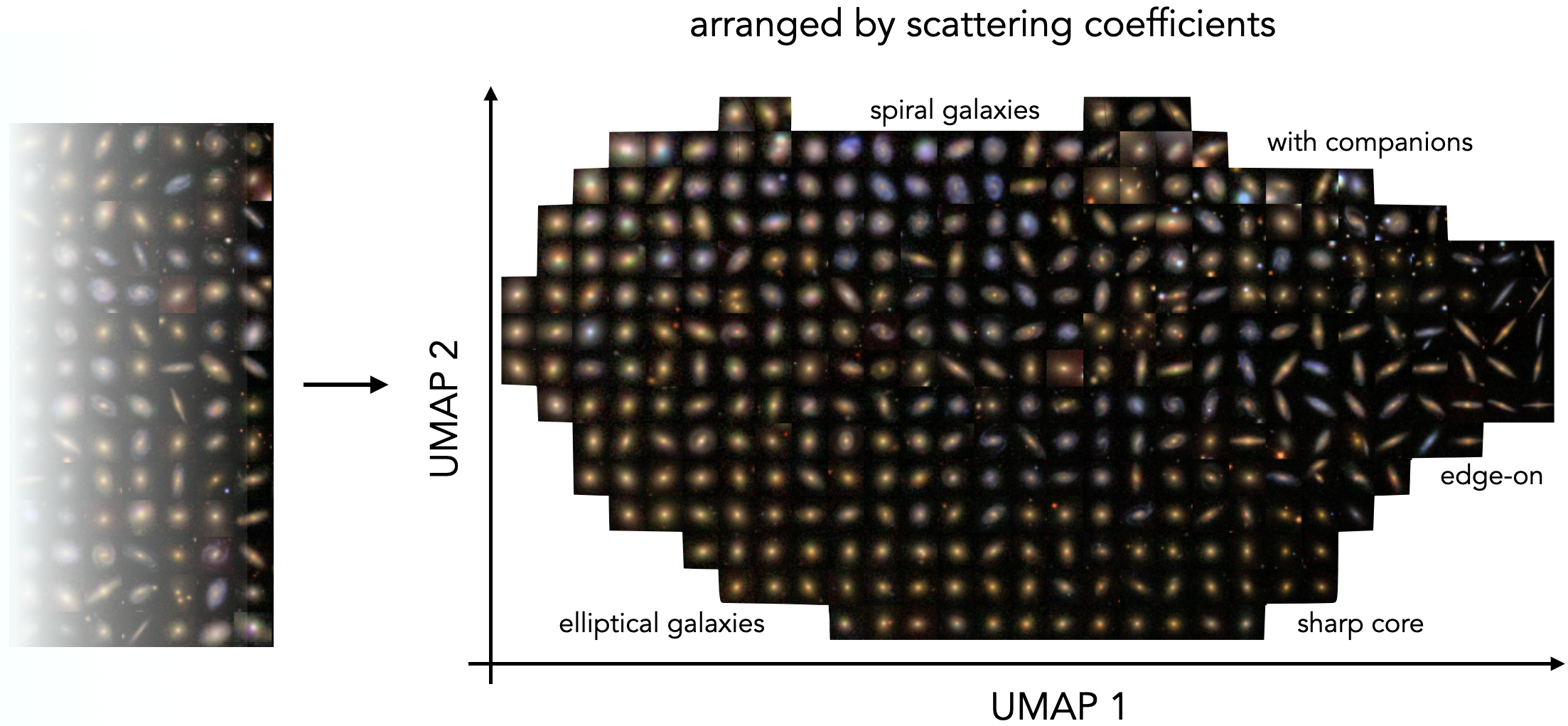}
    \caption{A gallery of randomly selected galaxy images (left) and a set arranged according to their 2nd-order scattering coefficients (right). A UMAP projection has been used to project the 30-dimensional representation onto a 2D plane.}
    \label{fig:galaxy}
\end{figure*}

\subsubsection*{Exploring objects}

Throughout this paper, we have primarily focused on textures, i.e. stationary ergodic fields. It is now interesting to mention that the scattering transform can also be useful outside of this context and can be used to characterize objects. To illustrate this point, as an example, we use a set of galaxy images and show its ability to organized them depending on their morphology.

The left panel of figure~\ref{fig:galaxy} shows a gallery of galaxy images randomly selected from the database of Sloan Digital Sky Survey \citep{York_2000}. It exhibits a variety of galaxies with different morphology and color. To organize them, we first compute the reduced second-order scattering coefficients (in this case 30) for each image (eq.~\ref{eq:s1s2}). These coefficients extract valuable morphological information expressed in a space of 30 dimensions. In order to visualize the overall distribution, we use UMAP \citep[Uniform Manifold Approximation and Projection,][]{McInnes_2018} to project these coefficients onto two dimensions and create a map. As can be seen, galaxies are automatically organized in this space based on their morphological properties. This, again, illustrates the expressiveness of the scattering coefficients to describe morphological properties. In summary, the scattering transform is a useful tool to perform exploratory data analysis with both stochastic fields and objects.

\subsubsection*{Model parameter inference}

As we discussed earlier, for parameter inference, the informativeness, compactness and robustness of the statistical estimator are desirable properties, all found in the scattering transform. An example of parameter inference application can be found in cosmology where a physical model of the distribution of matter is available and its parameters need to be estimated from observations. In such context, the scattering transform has been shown to outperform traditional statistics and is on a par with CNNs \citep{Cheng_2020}. Another example is in quantum chemistry where state energy is to be inferred from molecular configurations. In this context, translation and permutation invariance is the key, and the scattering transform also proves to be one of the best descriptors \citep{Hirn_2017}.

\section{Conclusion}

%% Scientific data analysis is a special type of data analysis.
%% Similar to judicial decisions.

The complexity of the stochastic fields typically studied in many scientific disciplines tends to lie between simple Gaussian random fields and the elaborate systems found in the biological world whose complexity appears boundless. This intermediate regime calls for an appropriate data analysis approach which goes beyond a simple power spectrum estimate but does not necessarily motivate the use of deep learning.

Deep learning is certainly a powerful method to extract information from complex data but comes at a high cost: it requires large amounts of data and computation, and designing deep convolutional neural networks is a tedious task, mainly done empirically and prone to over-fitting problems.

When extracting information from data, it seems desirable to use a tool or an estimator whose expressivity is on par with the expected complexity. In this paper, we advocate for the use of the scattering transform to extract information from many of the fields encountered in scientific analyses. This estimator, introduced by \cite{Mallat_2012} and \cite{Bruna_2013} in the mathematics and signal processing literatures, provides an approach to data analysis that, in many ways, conveniently stands \emph{in between} the power spectrum and CNNs. Remarkably, its expressivity can encompass a wide diversity of scientific processes, as illustrated in figure~\ref{fig:synthesis}. The scattering transform is useful not only for classification tasks but for the full spectrum of data analysis problems, from exploratory data analysis all the way to precise regression problems when models are available, i.e. two types of situations commonly met by scientists. In some cases, its performance is found to be on par with that of CNNs.

The scattering transform possesses a number of attractive properties, ideally suited for the analysis of textures (stationary ergodic fields) but also useful for localized objects:
\begin{itemize}
\item it is invariant to translation (and possibly rotation) and it preserves energy.
\item It is capable of extracting non-Gaussian or morphological information. It does so with an exponentially fast convergence for a wide range of fields. In many cases, a second-order only scattering transform is sufficient to extract the relevant morphological information.
\item It produces a compact set of informative and robust summary statistics with asymptotic normality. It is also stable to small deformations, a desirable property for both classification and parameter inference. 
\item Last but not least, its coefficients are interpretable. We showed that two sets of scattering coefficients are particularly informative:
\begin{itemize}
    \item a sparsity estimate $s_{21}$ which indicates whether fluctuations or structures are concentrated at certain positions or widely spread.
    \item a shape estimate $s_{22}$ (defined for two-dimensional fields) which describes the level of straight versus curved features present in the data.
\end{itemize}
\end{itemize}
As already shown in a number of scientific applications, the performance of the scattering transform can be on par with that of CNNs but allows one to use a well-defined estimator, independent of any training set and using a set of interpretable summary statistics, all important properties in scientific research. We believe that this approach can greatly benefit a wide range of scientific data analyses. In addition, understanding the core operations of the scattering transform allows one to decipher many key aspects of the inner workings of CNNs. 

\section*{Acknowledgements}

We thank St\'ephane Mallat for insightful discussions and feedback on the manuscript. This work was supported by the Packard Foundation and the generosity of Eric and Wendy Schmidt by recommendation of the Schmidt Futures program.

\bibliographystyle{mnras}
\bibliography{ST_theory}

\begin{thebibliography}{}
\makeatletter
\relax
\def\mn@urlcharsother{\let\do\@makeother \do\$\do\&\do\#\do\^\do\_\do\%\do\~}
\def\mn@doi{\begingroup\mn@urlcharsother \@ifnextchar [ {\mn@doi@}
  {\mn@doi@[]}}
\def\mn@doi@[#1]#2{\def\@tempa{#1}\ifx\@tempa\@empty \href
  {http://dx.doi.org/#2} {doi:#2}\else \href {http://dx.doi.org/#2} {#1}\fi
  \endgroup}
\def\mn@eprint#1#2{\mn@eprint@#1:#2::\@nil}
\def\mn@eprint@arXiv#1{\href {http://arxiv.org/abs/#1} {{\tt arXiv:#1}}}
\def\mn@eprint@dblp#1{\href {http://dblp.uni-trier.de/rec/bibtex/#1.xml}
  {dblp:#1}}
\def\mn@eprint@#1:#2:#3:#4\@nil{\def\@tempa {#1}\def\@tempb {#2}\def\@tempc
  {#3}\ifx \@tempc \@empty \let \@tempc \@tempb \let \@tempb \@tempa \fi \ifx
  \@tempb \@empty \def\@tempb {arXiv}\fi \@ifundefined
  {mn@eprint@\@tempb}{\@tempb:\@tempc}{\expandafter \expandafter \csname
  mn@eprint@\@tempb\endcsname \expandafter{\@tempc}}}

\bibitem[\protect\citeauthoryear{{Allys}, {Levrier}, {Zhang}, {Colling},
  {Regaldo-Saint Blancard}, {Boulanger}, {Hennebelle}  \& {Mallat}}{{Allys}
  et~al.}{2019}]{Allys_2019}
{Allys} E.,  {Levrier} F.,  {Zhang} S.,  {Colling} C.,  {Regaldo-Saint
  Blancard} B.,  {Boulanger} F.,  {Hennebelle} P.,   {Mallat} S.,  2019,
  \mn@doi [\aap] {10.1051/0004-6361/201834975}, \href
  {https://ui.adsabs.harvard.edu/abs/2019A&A...629A.115A} {629, A115}

\bibitem[\protect\citeauthoryear{{Allys}, {Marchand}, {Cardoso},
  {Villaescusa-Navarro}, {Ho}  \& {Mallat}}{{Allys} et~al.}{2020}]{Allys_2020}
{Allys} E.,  {Marchand} T.,  {Cardoso} J.~F.,  {Villaescusa-Navarro} F.,  {Ho}
  S.,   {Mallat} S.,  2020, \mn@doi [\prd] {10.1103/PhysRevD.102.103506}, \href
  {https://ui.adsabs.harvard.edu/abs/2020PhRvD.102j3506A} {102, 103506}

\bibitem[\protect\citeauthoryear{{And\'en} \& {Mallat}}{{And\'en} \&
  {Mallat}}{2011}]{AndenMallat_2011}
{And\'en} J.,  {Mallat} S.,  2011, International Society for Music Information
  Retrieval Conference, \href
  {https://www.di.ens.fr/data/publications/papers/ismir-final.pdf} {pp
  657--662}

\bibitem[\protect\citeauthoryear{{And\'en} \& {Mallat}}{{And\'en} \&
  {Mallat}}{2014}]{AndenMallat_2014}
{And\'en} J.,  {Mallat} S.,  2014, \mn@doi [IEEE Transactions on Signal
  Processing] {10.1109/TSP.2014.2326991}, \href
  {https://arxiv.org/abs/1304.6763} {62, 4114}

\bibitem[\protect\citeauthoryear{{Andreux} et~al.,}{{Andreux}
  et~al.}{2018}]{Kymatio_2018}
{Andreux} M.,  et~al., 2018, arXiv e-prints, \href
  {https://ui.adsabs.harvard.edu/abs/2018arXiv181211214A} {p. arXiv:1812.11214}

\bibitem[\protect\citeauthoryear{{Brochard}, {B{\l}aszczyszyn}, {Mallat}  \&
  {Zhang}}{{Brochard} et~al.}{2020}]{Brochard_2020}
{Brochard} A.,  {B{\l}aszczyszyn} B.,  {Mallat} S.,   {Zhang} S.,  2020, arXiv
  e-prints, \href {https://arxiv.org/abs/2010.14928} {p. arXiv:2010.14928}

\bibitem[\protect\citeauthoryear{{Bruna} \& {Mallat}}{{Bruna} \&
  {Mallat}}{2013}]{Bruna_2013}
{Bruna} J.,  {Mallat} S.,  2013, \mn@doi [IEEE Transactions on Pattern Analysis
  and Machine Intelligence] {doi: 10.1109/TPAMI.2012.230}, \href
  {https://arxiv.org/abs/1203.1513} {35, 1872}

\bibitem[\protect\citeauthoryear{Bruna \& Mallat}{Bruna \&
  Mallat}{2019}]{bruna2019multiscale}
Bruna J.,  Mallat S.,  2019, Mathematical Statistics and Learning, \href
  {https://arxiv.org/abs/1801.02013} {1, 257}

\bibitem[\protect\citeauthoryear{{Bruna}, {Mallat}, {Bacry}  \& J.-F.}{{Bruna}
  et~al.}{2015}]{Bruna_2015}
{Bruna} J.,  {Mallat} S.,  {Bacry} E.,   J.-F. M.,  2015, \mn@doi [The Annals
  of Statistics] {10.1214/14-AOS1276}, \href {https://arxiv.org/abs/1311.4104}
  {43, 323}

\bibitem[\protect\citeauthoryear{{Carron}}{{Carron}}{2011}]{Carron_2011}
{Carron} J.,  2011, \mn@doi [\apj] {10.1088/0004-637X/738/1/86}, \href
  {https://ui.adsabs.harvard.edu/abs/2011ApJ...738...86C} {738, 86}

\bibitem[\protect\citeauthoryear{{Cheng} \& {M{\'e}nard}}{{Cheng} \&
  {M{\'e}nard}}{2021}]{Cheng_2021}
{Cheng} S.,  {M{\'e}nard} B.,  2021, \mn@doi [\mnras] {10.1093/mnras/stab2102},
  \href {https://ui.adsabs.harvard.edu/abs/2021MNRAS.507.1012C} {507, 1012}

\bibitem[\protect\citeauthoryear{{Cheng}, {Ting}, {M{\'e}nard}  \&
  {Bruna}}{{Cheng} et~al.}{2020}]{Cheng_2020}
{Cheng} S.,  {Ting} Y.-S.,  {M{\'e}nard} B.,   {Bruna} J.,  2020, \mn@doi
  [\mnras] {10.1093/mnras/staa3165}, \href
  {https://ui.adsabs.harvard.edu/abs/2020MNRAS.tmp.2970C} {}

\bibitem[\protect\citeauthoryear{{Eickenberg}, {Exarchakis}, {Hirn}, {Mallat}
  \& {Thiry}}{{Eickenberg} et~al.}{2018}]{Eickenberg_2018}
{Eickenberg} M.,  {Exarchakis} G.,  {Hirn} M.,  {Mallat} S.,   {Thiry} L.,
  2018, \mn@doi [\jcp] {10.1063/1.5023798}, \href
  {https://ui.adsabs.harvard.edu/abs/2018JChPh.148x1732E} {148, 241732}

\bibitem[\protect\citeauthoryear{{Glinsky} et~al.,}{{Glinsky}
  et~al.}{2020}]{Glinsky_2020}
{Glinsky} M.~E.,  et~al., 2020, \mn@doi [Physics of Plasmas]
  {10.1063/5.0010781}, \href {https://arxiv.org/abs/2005.01600} {27, 112703}

\bibitem[\protect\citeauthoryear{{Guth}, {Zarka}  \& {Mallat}}{{Guth}
  et~al.}{2021}]{Guth_2021}
{Guth} F.,  {Zarka} J.,   {Mallat} S.,  2021, arXiv e-prints, \href
  {https://arxiv.org/abs/2110.05283} {p. arXiv:2110.05283}

\bibitem[\protect\citeauthoryear{{Hirn}, {Mallat}  \& {Poilvert}}{{Hirn}
  et~al.}{2017}]{Hirn_2017}
{Hirn} M.,  {Mallat} S.,   {Poilvert} N.,  2017, \mn@doi [Multiscale Modeling
  \& Simulation] {10.1137/16M1075454}, \href
  {https://ui.adsabs.harvard.edu/abs/2016arXiv160504654H} {15, 827–863}

\bibitem[\protect\citeauthoryear{{Hubel} \& {Wiesel}}{{Hubel} \&
  {Wiesel}}{1968}]{Hubel_1968}
{Hubel} D.,  {Wiesel} T.,  1968, \mn@doi [The Journal of physiology]
  {10.1113/jphysiol.1968.sp008455}, 195, 215

\bibitem[\protect\citeauthoryear{{Kavalerov}, {Li}, {Czaja}  \&
  {Chellappa}}{{Kavalerov} et~al.}{2019}]{Kavalerov_2019}
{Kavalerov} I.,  {Li} W.,  {Czaja} W.,   {Chellappa} R.,  2019, arXiv e-prints,
  \href {https://ui.adsabs.harvard.edu/abs/2019arXiv190606804K} {p.
  arXiv:1906.06804}

\bibitem[\protect\citeauthoryear{Krizhevsky, Sutskever  \& Hinton}{Krizhevsky
  et~al.}{2012}]{Krizhevsky_2012}
Krizhevsky A.,  Sutskever I.,   Hinton G.~E.,  2012, in Pereira F.,  Burges C.
  J.~C.,  Bottou L.,   Weinberger K.~Q.,  eds, ~ Vol. 25, Advances in Neural
  Information Processing Systems. Curran Associates, Inc., \url
  {https://proceedings.neurips.cc/paper/2012/file/c399862d3b9d6b76c8436e924a68c45b-Paper.pdf}

\bibitem[\protect\citeauthoryear{{Mallat}}{{Mallat}}{2009}]{Mallat_2009}
{Mallat} S.,  2009, {A wavelet tour of signal processing. The Sparse Way.},
  third edn.
Academic Press, Boston, \mn@doi{10.1016/B978-0-12-374370-1.X0001-8}

\bibitem[\protect\citeauthoryear{{Mallat}}{{Mallat}}{2012}]{Mallat_2012}
{Mallat} S.,  2012, \mn@doi [Communications on Pure and Applied Mathematics]
  {10.1002/cpa.21413}, \href {https://arxiv.org/abs/1101.2286} {65, 1331}

\bibitem[\protect\citeauthoryear{{Mallat}, {Zhang}  \& {Rochette}}{{Mallat}
  et~al.}{2019}]{Mallat_2019}
{Mallat} S.,  {Zhang} S.,   {Rochette} G.,  2019, \mn@doi [Information and
  Inference: A Journal of the IMA] {10.1093/imaiai/iaz019}, \href
  {https://arxiv.org/abs/arXiv:1810.12136} {9, 721}

\bibitem[\protect\citeauthoryear{{McInnes}, {Healy}  \& {Melville}}{{McInnes}
  et~al.}{2018}]{McInnes_2018}
{McInnes} L.,  {Healy} J.,   {Melville} J.,  2018, arXiv e-prints, \href
  {https://ui.adsabs.harvard.edu/abs/2018arXiv180203426M} {p. arXiv:1802.03426}

\bibitem[\protect\citeauthoryear{{Olshausen} \& {Field}}{{Olshausen} \&
  {Field}}{1996}]{Olshausen_1996}
{Olshausen} B.,  {Field} D.,  1996, \mn@doi [Nature] {10.1038/381607a0}, 381,
  607

\bibitem[\protect\citeauthoryear{{Portilla} \& {Simoncelli}}{{Portilla} \&
  {Simoncelli}}{2000}]{Portilla_2000}
{Portilla} J.,  {Simoncelli} E.,  2000, \mn@doi [International Journal of
  Computer Vision] {10.1023/A:1026553619983}, 40, 49–70

\bibitem[\protect\citeauthoryear{Prochaska, Cornillon  \& Reiman}{Prochaska
  et~al.}{2021}]{Prochaska_2021}
Prochaska J.~X.,  Cornillon P.~C.,   Reiman D.~M.,  2021, \mn@doi [Remote
  Sensing] {10.3390/rs13040744}, 13

\bibitem[\protect\citeauthoryear{{Regaldo-Saint Blancard}, {Levrier}, {Allys},
  {Bellomi}  \& {Boulanger}}{{Regaldo-Saint Blancard}
  et~al.}{2020}]{Blancard_2020}
{Regaldo-Saint Blancard} B.,  {Levrier} F.,  {Allys} E.,  {Bellomi} E.,
  {Boulanger} F.,  2020, \mn@doi [\aap] {10.1051/0004-6361/202038044}, \href
  {https://ui.adsabs.harvard.edu/abs/2020A&A...642A.217R} {642, A217}

\bibitem[\protect\citeauthoryear{{Regaldo-Saint Blancard}, {Allys},
  {Boulanger}, {Levrier}  \& {Jeffrey}}{{Regaldo-Saint Blancard}
  et~al.}{2021}]{Blancard_2021}
{Regaldo-Saint Blancard} B.,  {Allys} E.,  {Boulanger} F.,  {Levrier} F.,
  {Jeffrey} N.,  2021, \mn@doi [\aap] {10.1051/0004-6361/202140503}, \href
  {https://ui.adsabs.harvard.edu/abs/2021A&A...649L..18R} {649, L18}

\bibitem[\protect\citeauthoryear{{Saydjari}, {Portillo}, {Slepian}, {Kahraman},
  {Burkhart}  \& {Finkbeiner}}{{Saydjari} et~al.}{2021}]{Saydjari_2021}
{Saydjari} A.~K.,  {Portillo} S. K.~N.,  {Slepian} Z.,  {Kahraman} S.,
  {Burkhart} B.,   {Finkbeiner} D.~P.,  2021, \mn@doi [\apj]
  {10.3847/1538-4357/abe46d}, \href
  {https://ui.adsabs.harvard.edu/abs/2021ApJ...910..122S} {910, 122}

\bibitem[\protect\citeauthoryear{{Sifre} \& {Mallat}}{{Sifre} \&
  {Mallat}}{2013}]{Sifre_2013}
{Sifre} L.,  {Mallat} S.,  2013, the IEEE Conference on Computer Vision and
  Pattern Recognition (CVPR), \href
  {https://openaccess.thecvf.com/content_cvpr_2013/html/Sifre_Rotation_Scaling_and_2013_CVPR_paper.html}
  {pp 1233--1240}

\bibitem[\protect\citeauthoryear{{Sinz}, {Swift}, {Brumwell}, {Liu}, {Kim},
  {Qi}  \& {Hirn}}{{Sinz} et~al.}{2020}]{Sinz_2020}
{Sinz} P.,  {Swift} M.~W.,  {Brumwell} X.,  {Liu} J.,  {Kim} K.~J.,  {Qi} Y.,
  {Hirn} M.,  2020, \mn@doi [{\jcp}] {10.1063/5.0016020}, \href
  {https://arxiv.org/abs/2006.01247} {153, 084109}

\bibitem[\protect\citeauthoryear{{Szegedy}, {Zaremba}, {Sutskever}, {Bruna},
  {Erhan}, {Goodfellow}  \& {Fergus}}{{Szegedy} et~al.}{2013}]{Szegedy_2013}
{Szegedy} C.,  {Zaremba} W.,  {Sutskever} I.,  {Bruna} J.,  {Erhan} D.,
  {Goodfellow} I.,   {Fergus} R.,  2013, arXiv e-prints, \href
  {https://arxiv.org/abs/1312.6199} {p. arXiv:1312.6199}

\bibitem[\protect\citeauthoryear{{York} et~al.,}{{York}
  et~al.}{2000}]{York_2000}
{York} D.~G.,  et~al., 2000, \mn@doi [\aj] {10.1086/301513}, \href
  {https://ui.adsabs.harvard.edu/abs/2000AJ....120.1579Y} {120, 1579}

\bibitem[\protect\citeauthoryear{{Zarka}, {Thiry}, {Angles}  \&
  {Mallat}}{{Zarka} et~al.}{2019}]{Zarka_2019}
{Zarka} J.,  {Thiry} L.,  {Angles} T.,   {Mallat} S.,  2019, arXiv e-prints,
  \href {https://arxiv.org/abs/1910.03561} {p. arXiv:1910.03561}

\bibitem[\protect\citeauthoryear{{Zarka}, {Guth}  \& {Mallat}}{{Zarka}
  et~al.}{2020}]{Zarka_2020}
{Zarka} J.,  {Guth} F.,   {Mallat} S.,  2020, arXiv e-prints, \href
  {https://arxiv.org/abs/2012.10424} {p. arXiv:2012.10424}

\bibitem[\protect\citeauthoryear{{Zhang} \& {Mallat}}{{Zhang} \&
  {Mallat}}{2021}]{Zhang_2021}
{Zhang} S.,  {Mallat} S.,  2021, \mn@doi [Applied and Computational Harmonic
  Analysis] {https://doi.org/10.1016/j.acha.2021.01.003}, \href
  {https://arxiv.org/abs/1911.10017} {53, 199}

\makeatother
\end{thebibliography}

\appendix

\section{Generative models}
\label{app:synthesis}

When it comes to non-Gaussian fields, there is no universal metric to compare different statistical estimators. In order to show the ability of the scattering transform to characterise a field, it is informative to build a generative model and assess its performance.

A generative model can visualise what an input field looks like in the eye of a summary statistics. It randomly generates new fields with the same summary statistics as the input field, which form a `microcanonical ensemble' of fields given the particular values of the statistics \citep[see, e.g.,][]{Portilla_2000, bruna2019multiscale}. If the generated fields have similar textures to the input field, or in other words, if the input field is a representative and typical one in the ensemble, then it means that the statistics capture the main features of the input field.

As shown in figure~\ref{fig:synthesis}, fields generated with the power spectrum statistics are Gaussian random field. The synthesis results with additional bispectrum information (the 3-point statistics in Fourier space) provide an improvement but are still far from ideal. In contrast,  we show the field generation results using the scattering statistics for different fields chosen from various disciplines of physics. The striking similarity between the input and generated fields is  evidence of the power of the scattering statistics to characterise various realistic non-Gaussian textures in physics.

To build these field realizations, we starting from a random field and then gradually modify it in order to minimise the difference between its summary statistics and that of the input field. We used the `adam' optimiser in the python package \texttt{torch.optim} to implement the minimisation. Codes for the corresponding figures are available at \href{https://github.com/SihaoCheng/scattering_transform}{https://github.com/SihaoCheng/scattering\_transform}.

There are several caveats about such generative models to keep in mind. For example, the image quality somewhat depends on the optimisation's initial condition and the choice of loss function. More importantly, the concept of information must always be related to a particular task. In the image generation case, this task is `to distinguish fields under the metric of human eyes and brains', which can be different from the task of `inferring the underlying physical parameters'. Nevertheless, in practice, the generation and physical inference abilities are often closely relative, which may interestingly suggest that the human visual system is well-evolved and optimised to extract information from the physical world.

\section{Appropriate wavelets}
\label{sec:wavelets}

\begin{figure}
    \centering
    \includegraphics[width=\columnwidth]{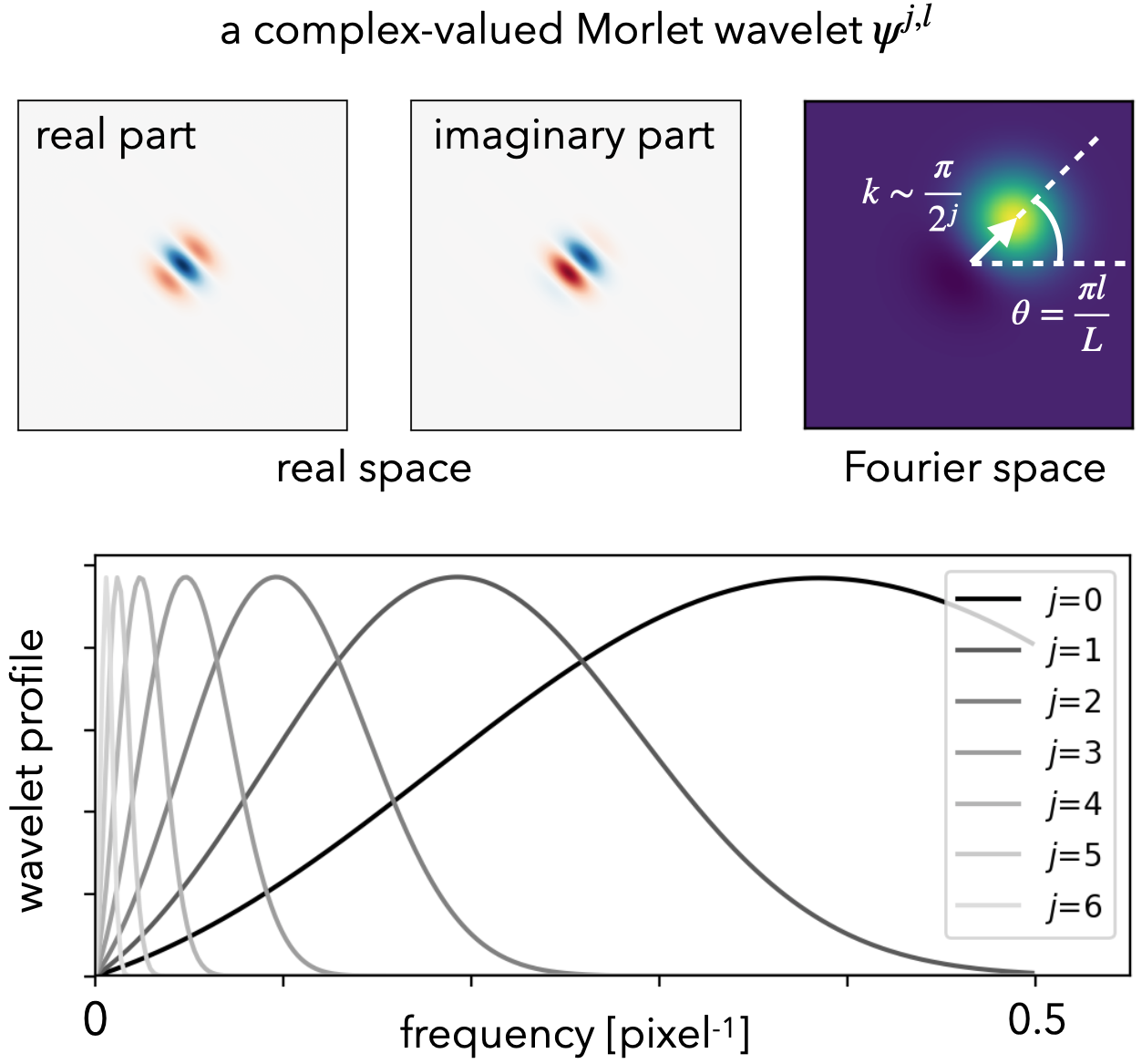}
    \caption{The profile of Morlet wavelet(s) in real and Fourier space.}
    \label{fig:Gabor}
\end{figure}

Wavelets are localized oscillations in real space and band-pass windows in Fourier space. A family of wavelet is composed of wavelets with the same shape but different sizes (or orientations). The choice of the wavelets used with the scattering transform however needs to satisfy a number of criteria in order to be `admissible' and lead to all the desired stability and conservation properties listed in section~\ref{sec:properties}. 

First, the scattering transform makes use of complex wavelets which provides a convenient modulus/phase separation and which can be used to transform fluctuations in a given field towards lower frequencies. To use all the input information and conserve energy, the wavelets need to homogeneously cover the full extend of the Fourier space, except at zero frequency. 
In addition, in order to be stable to small deformations their bandwidth need to be comparable to their central frequency. 

A family of wavelets can be constructed by dilating and rotating a \emph{mother} wavelet $\psi(x)$. A common choice is to use dyadic wavelets, meaning the dilation factor is two. In general, this choice provides a good compromise between the need for separating scales and probing scale interactions in the scattering transform. In one dimension cases, we can define these multiscale dyadic wavelets $\psi^{j,l}(x)$ for any scale index $j \in \mathbb{Z}$ by
$$
\psi^{j,l}(x)=\frac{1}{2^{j}} \psi\left(\frac{x}{2^j}\right)\;.
$$
In two dimensions, where both scales and orientations have to be considered, we can add a set of rotation operations $r_l$ corresponding to rotation angles $\pi l / L$, where $L$ is a pre-determined integer and $l$ is an integer between 0 and $L$
$$
\psi^{j,l}(\vec{x})=\frac{1}{(2^{j})^2} \psi\left(\frac{r_l^{-1} \vec{x}}{2^j}\right)\;.
$$
Its Fourier transform is $\hat{\psi}^{j,l}(\vec{k})=\hat{\psi}(r_l^{-1}\vec{k}/2^{-j})$. So, if the mother wavelet has a central frequency at $k_0$, then $\hat{\psi}^{j}$ has a support centered at $2^{-j} k_0$ and a bandwidth proportional to $2^{-j}$. Also, these wavelets have the same height in Fourier space. An illustration is shown in figure~\ref{fig:Gabor}. 

For wavelets, imposing that the Fourier domain is covered by the filters $\psi$ without holes is called the Littlewood-Paley inequality. It guarantees that the wavelet is invertible and stable, with bounds which depend upon $\eta$ \citep{Mallat_2009}
\begin{equation}
    (1-\eta)^{2} \leqslant  \frac{1}{2}\,\sum_{j,l}\left|\widehat{\psi}^{j,l}(k)\right|^{2} \leqslant(1+\eta)^{2}
\end{equation}
Different mother wavelets can be chosen. They can lead to non-orthogonal and redundant wavelets $\psi^{j,l}(x)$ which will satisfy the above relation with different bounds. 

A simple choice is the Gabor wavelet, which is a Gaussian window in Fourier space and a Fourier mode modulated by a Gaussian envelope in real space. The advantage of using Gaussian profiles is that they are sufficiently localized in both Fourier and real spaces. In arbitrary dimensions, it can be written as
 \begin{eqnarray}
    \tilde{G}(\bm{k}) &=& e^{-(\bm{k}-\bm{k}_0)^T\bm{\Sigma}(\bm{k}-\bm{k}_0)/2}\nonumber\\
    G(\bm{x}) &=& e^{-\bm{x}^T\bm{\Sigma}^{-1}\bm{x}/2}\,e^{i \bm{k}_0 \cdot \bm{x}} / \sqrt{|\bm{\Sigma}|}
    % \frac{1}{\sqrt{|\bm{\Sigma}|}}
\end{eqnarray}
where $\bm{\Sigma}$ is the covariance matrix describing the size, shape, and orientation of the Gaussian envelope, and $\bm{k}_0$ determines the frequency of the oscillation. 
To obtain maximum rotational symmetry, usually $\bm{\Sigma}$ is selected to have only one eigen-value different from the others, and $\bm{k}_0$ to be along that eigen-direction. Thus we denote the eigen-value along $\bm{k}_0$ by $\sigma^2$ and the others by $\sigma^2/s^2$. The parameter $s$ is the ratio of transverse to radial width of the wavelet in Fourier space.
%
% Note that the product $k_0\sigma$ determines the number of oscillations within $\pm\pi\approx3$ standard deviation of the Gaussian envelope and allows for a trade off between spatial and frequency resolution.

However, the admissibility condition requires wavelets to be band-pass filters. Unfortunately, a Gaussian profile in Fourier space does not vanish at the origin. A simple solution is to introduce an offset $\beta$ before the Gaussian modulation. In Fourier space this is equivalent to subtracting another Gaussian profile centred at the origin to cancel out the zero-frequency component. Families of wavelets created in this way are called Morlet wavelets. Formally,
\begin{equation}
    \psi(\bm{x}) = \frac{1}{\sqrt{|\bm{\Sigma}|}}e^{-\bm{x}^T\bm{\Sigma}^{-1}\bm{x}/2}\left(e^{i \bm{k}_0 \cdot \bm{x}}-\beta\right)\,,
\end{equation}
where $\beta=e^{-\bm{k}_0^T\bm{\Sigma}\bm{k}_0/2}$. Its Fourier transform is
\begin{equation}
\tilde{\psi}(\bm{k}) = \tilde{G}(\bm{k}) - \beta e^{-\bm{k}^T\bm{\Sigma}\bm{k}/2}\,.
\end{equation}

In the two-dimension case, we follow the settings used in the `kymatio' package,
\begin{align}
\label{eq:Morlet_param}
    \sigma&=0.8\times2^j\;,~~~    k_0=\frac{3\pi}{4\times2^j}\;,~~~    s=4/L\,,
\end{align}
where $\sigma$ is in unit of pixels, $j$ is an integer starting from 0, and $k_0$ is always between 0 and 2$\pi$. This choice allow a family of Morlet wavelets best covers the whole Fourier space with a dyadic sequence of scales ($2^j$).
% Examples of the Morlet wavelets we use are shown in figure~\ref{fig:wavelets}. 
Within the wavelet envelope, there are about 2 cycles of oscillations, as $k_0\sigma\approx2$.

Strictly speaking, when sampled at discrete scales, general wavelet transforms only promise a rough conservation of `energy' within a finite error factor. 
% For example, the Morlet wavelets sampled at dyadic scales guarantees a difference smaller than about $<$5\,\%. 
To obtain strictly conservation, the wavelet profile should be properly chosen so that the square of their profile add up exactly to unity (including the zero frequency intensity) over the Fourier space.

\section{Scattering coefficients and N-point functions}
\label{app:S_to_Npoint}

\begin{figure}
    \centering
    \includegraphics[width=\columnwidth]{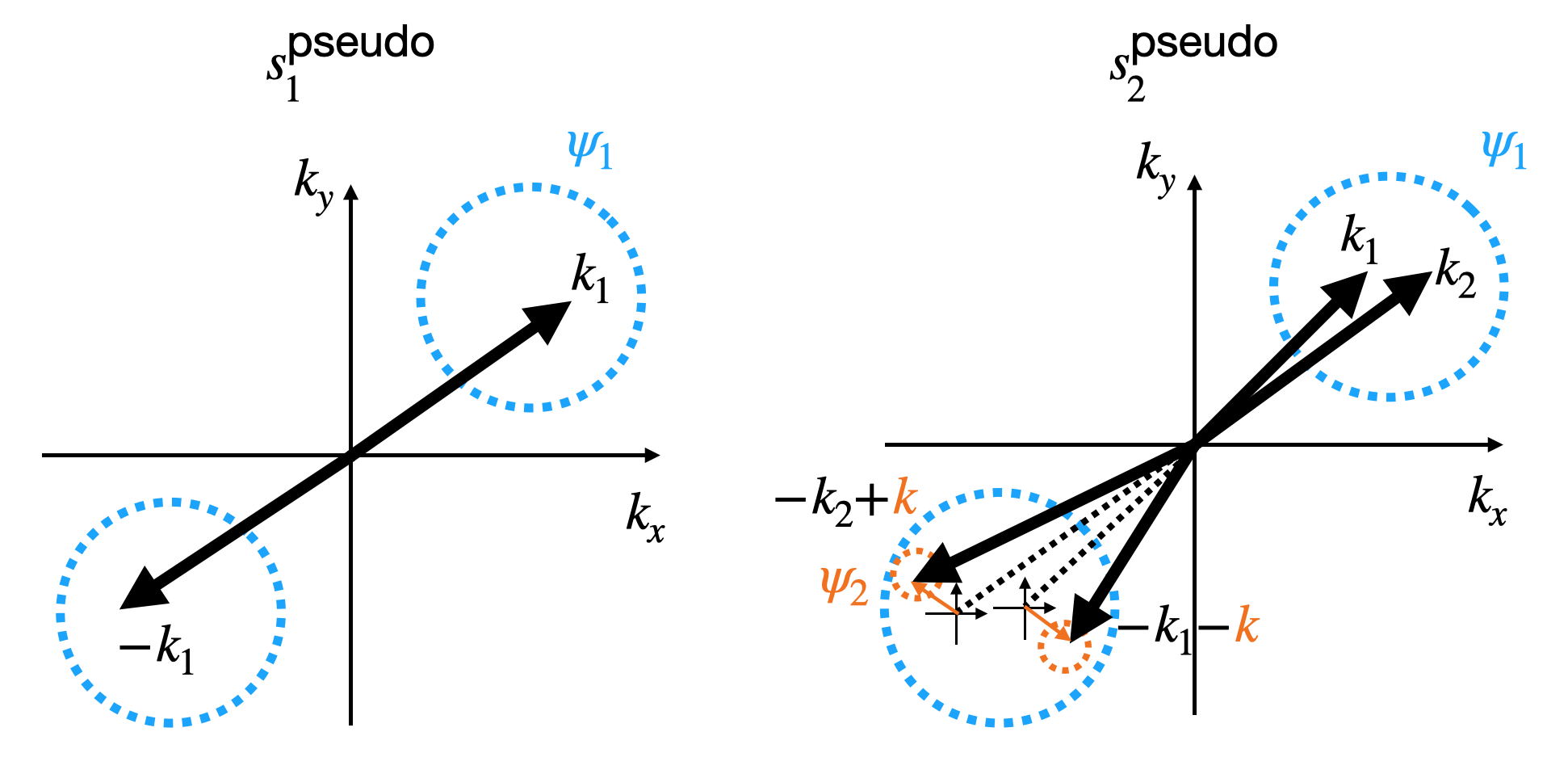}
    \caption{The 2-point and 4-point functions corresponding to scattering coefficients when replacing the modulus operation by modulus squared.}
    \label{fig:4point}
\end{figure}

The scattering transform is similar to $N$-point functions in several ways. Indeed, if we use the following replacement:
\begin{align}
    \text{modulus} \rightarrow \text{modulus squared,}
\end{align}
then the $n$th order scattering coefficients $S_n$ will become some binned 2$^n$-point functions. In other words, if one elevates the power of its non-linear operation, the scattering transform becomes exactly a high-order statistic. Let's call these power-2 scattering coefficients as ``pseudo scattering coefficients''. First, we note that the pseudo scattering coefficients at each order are defined from the previous order by
\begin{align}
    \nonumber S_{n+1}^{\text{pseudo}} &\equiv \langle I_{n+1}^{\text{pseudo}} \rangle\\
    \nonumber &\equiv \int |I_n\star\psi_{n+1}|^2 d\bm{x}\\
    \nonumber &= \int |\tilde{I_n}\cdot\tilde{\psi}_{n+1}|^2 d\bm{k}\\
    &= \int P_n(\bm{k})\cdot \tilde{\psi}_{n+1}(\bm{k})^2 d\bm{k}\,,
\end{align}
where $P_n(k)$ is the power spectrum of $I_n$. This equation shows that each pseudo scattering coefficient is an averaged power spectrum of the previous-order pseudo scattering field. The derivation makes use of three properties: (1) Fourier transform is orthonormal; (2) a convolution in real space is a multiplication in Fourier space; and (3) wavelets $\psi_{n+1}$ are real-valued band-pass filters in Fourier space.

Applying the above statement to the first order, we obtain that $S_1^{\text{pseudo}}$ are averaged (binned) power spectrum (2-point function) of the input field, weighted by wavelets:
\begin{align}
    \nonumber S_1^{\text{pseudo}} &= \int P_0(\bm{k})\cdot \tilde{\psi_1}(\bm{k})^2 d\bm{k}\\
    &= \text{ binned power spectrum of }I_0\,.
\end{align}
Applying it to the second order, $S_2^{\text{pseudo}}$ are binned power spectrum of the intermediate scattering fields $I_1^{\text{pseudo}}$, whose Fourier transform is
\begin{align}
    \nonumber \tilde{I}_1^{\text{pseudo}}(\bm{k}) &= \tilde{(I_0\star\psi_1)(I_0\star\psi_1)^*}\\
    &= \int \tilde{I_0}(\bm{k'}) \tilde{I_0}(\bm{k'}+\bm{k})^* \cdot [ \tilde{\psi_1}(\bm{k'}+\bm{k}) \tilde{\psi_1}(\bm{k'})] \cdot d\bm{k'}\,.
\end{align}
Again, because wavelets are real-valued in Fourier space ($\tilde{\psi}=\tilde{\psi}^*$), we have
\begin{align}
    \nonumber S_2^{\text{pseudo}} =& \int P_1\cdot \tilde{\psi_2}^2 d\bm{k}\\
    \nonumber =& \int \tilde{I_1^{\text{pseudo}}} \cdot \tilde{I_1^{\text{pseudo}}}^* \cdot \tilde{\psi_2}^2 d\bm{k}\\
    \nonumber =& \iiint \tilde{I_0}(\bm{k_1})\tilde{I_0}(\bm{k_1}+\bm{k})^* \tilde{I_0}(-\bm{k_2})^*\tilde{I_0}(-\bm{k_2}+\bm{k})\\
    \nonumber &\cdot \left[\tilde{\psi_1}(\bm{k_1}) \tilde{\psi_1}(\bm{k_1}+\bm{k})  \tilde{\psi_1}(-\bm{k_2}) \tilde{\psi_1}(-\bm{k_2}+\bm{k}) \tilde{\psi_2}(\bm{k})^2 \right]\\ \nonumber &\cdot d\bm{k_1} d\bm{k_2} d\bm{k} \\
    =& \text{ binned tri-spectrum of }I_0\,.
\end{align}
The integrand is the product of four Fourier coefficients of the input field. Each coefficient is weighted by the first wavelet $\tilde{\psi_1}$, and their difference $\bm{k}$ is weighted by the second wavelet $\tilde{\psi_2}$. An illustration of these 4-point configurations in Fourier space is shown in figure~\ref{fig:4point}.

Higher-order cases can be derived in a similar way, revealing that the $n$th-order pseudo scattering coefficients $S_2^{\text{pseudo}}$ are exactly some averaged $2^n$-point functions. Note that the ``binning'' or averaging weights are defined by the wavelets $\psi_1$, $\psi_2$, ... $\psi_n$ used at each order, so they are not equivalent to any squeezed limit in either real or Fourier domain.

\section{Low-order statistics: Don't amplify the tail}
\label{app:folding}

The scattering transform uses successive modulus operations instead of multiplications as it non-linearities. To compare the two approaches, we use a single example with one random variable instead of a field with infinite variables, as a toy model. Let us assume that this variable $x$ follows a probability distribution $p(x)$. The insights gained here will also apply to random fields.

\begin{figure}
    \centering
    \includegraphics[width=\columnwidth]{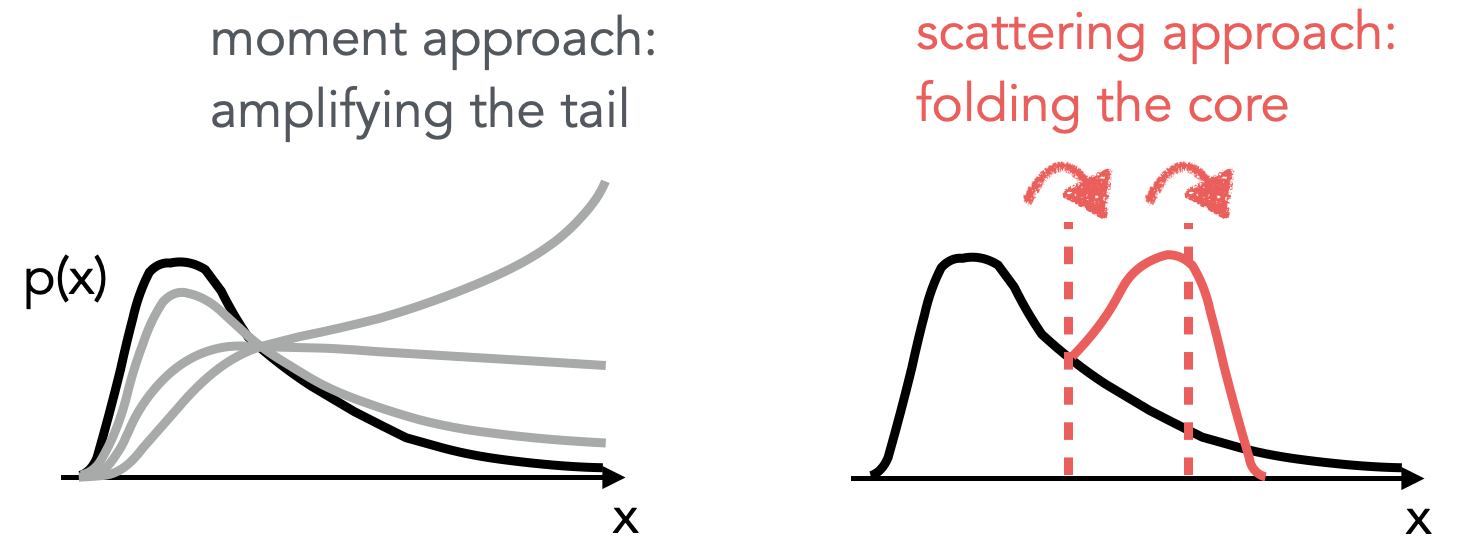}
    \caption{Comparison between the moment approach with scattering approach: calculating moments is equivalent to amplifying the tail of $p(x)$ and then integrating over $x$. When the tail is heavy, moments can divergence. In contrast, the scattering approach is equivalent to folding the core of $p(x)$.
    }
    \label{fig:ST_folding1}
\end{figure}

In this over-simplified case, the wavelet convolutions are mathematically reduced to a simple subtraction of the mean, and then the successive modulus can be described as folding the core of the distribution, as illustrated in figure~\ref{fig:ST_folding1}. Formally, the scattering transform can be defined by first repeatedly transforming a sample of the random variable $x$ through subtracting the sample mean and taking absolute value $x_n\rightarrow x_{n+1} \equiv |x_n - \bar{x}_n|$.
% \begin{align}
% \label{eq:1D_scattering}
%     x_0 &= \text{original sample }\nonumber\\
%     x_1 &= |x_0 - \bar{x}_0|\nonumber\\
%     x_2 &= |x_1 - \bar{x}_1|\\
%     &...\nonumber
% \end{align}
One can then take the mean of each transformed set $S_n \equiv \bar{x}_n$ as the $n$th-order `scattering coefficients' in this single-variable case. Similar to eq.~\ref{eq:energy_partition}, these statistics provide a partition of the sample variance, which indicates stability and robustness.

% \footnote{With a finite sample drawn from a distribution, there is a limit of how well one can constrain the distribution parameters, formalised by the Fisher information matrix and Cram\'er-Rao inequality. In practice, one usually uses summary statistics instead of the original sample data points to constrain distribution parameters (an idea similar to what is described in section~\ref{sec:characterisation}). The question here is, can these summary statistics squeeze out all available information and reach the limit set by the Fisher information matrix?} 

\begin{figure}
    \centering
    \includegraphics[width=0.9\columnwidth]{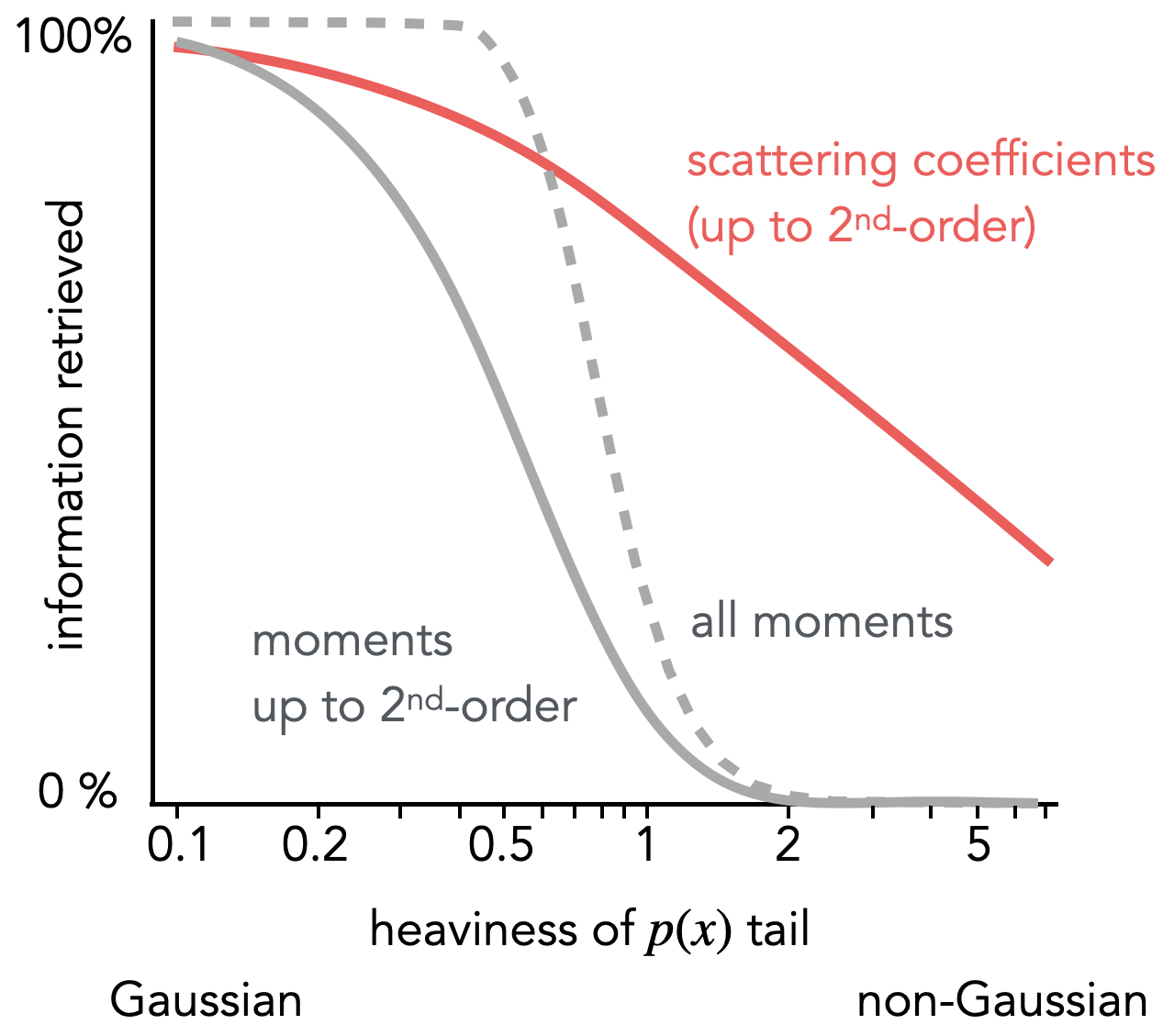}
    \caption{The fraction of Fisher information contained in moment and scattering statistics, respectively, for log-normal distributions with varying tails. The x axis is defined as $\sigma_\delta \equiv \sqrt{\exp(\sigma^2) - 1}$, where $\sigma$ is the variance of the corresponding Gaussian variable, and $\sigma_\delta$ is also the ratio of the standard deviation to the mean of the log-normal distribution. The folding strategy of the scattering transform works much better in the heavy-tail regime than traditional moments.
    }
    \label{fig:ST_folding2}
\end{figure}

Moment expansions seem to lure some in hoping that information about a probability distribution function can be extracted through its series of moments with arbitrarily high precision. Unfortunately, this approach can fail in some cases, even if all higher-order moments are taken into account. This problematic regime is met when the distribution is heavy tailed, which is the case of a wide range of physical fields of interest.
To explore this more quantitatively, one can compare the amount of information extracted by the moment approach as quantified by the Fisher information and compared to the theoretical limit given by the Cram\'er-Rao inequality. \citet{Carron_2011} performed such an experiment using a family of log-normal distributions $p(x|\mu, \sigma)=\exp[-(\ln x-\mu)^2/2\sigma^2] / (x\sigma\sqrt{2\pi})$, with varying $\sigma$ to induce different levels of non-Gaussianities. As shown in figure~\ref{fig:ST_folding2}, even if all the higher-order moments are accessible, the amount of information one can extract decreases rapidly with the heaviness of the tail. In fact, for any distributions with a tail decaying slower than exponential, the series of moments (even infinite) no longer forms a complete set. In practice, it is also important to point out that the amount of information captured by the leading orders is quickly diluted. 
% \citep[see figure 2 of][]{Carron_2011}.
What about the scattering approach? In figure~\ref{fig:ST_folding2}, we show results of the same experiment with the one-variable scattering coefficients. As can be seen, the ability to extract information with heavy-tailed distributions is substantially higher and the overall decay much slower.
% These repeated centered-modulus statistics retrieve much more information in the highly non-Gaussian regime 
% \footnote{Different from \citet{Carron_2011}, we obtained the derivatives in Fisher matrix numerically instead of using analytical expressions, as no such expressions exist for the scattering statistics yet. However, we did validate our method by applying it to moments and confirmed its consistency with the analytical method. 

% There is another difference between moments and repeated modulus statistics. The repeated modulus statistics is similar to the central moments such as the variance: the higher-order ones are defined from sample statistics of the lower-order ones. As a result, the expectation of sample statistics may depend on sample size. However, this is in general not a worry for the scattering coefficients calculated from a field, because the analogue to `sample size', the number of pixels, is usually huge.}.

\bsp	% typesetting comment
\label{lastpage}
\end{document}